\documentclass[pra,superscriptaddress,amsmath,amssymb,twocolumn]{revtex4-2}
\usepackage{graphicx}
\usepackage{subfigure}
\usepackage{adjustbox}
\usepackage{bm}
\usepackage{bbm}
\usepackage{color}
\usepackage{braket}
\usepackage{standalone}
\usepackage{multirow}
\usepackage{tikz}
\usepackage{mathrsfs}
\usepackage{dsfont}
\usepackage[colorlinks,bookmarks=true,citecolor=blue,linkcolor=blue,urlcolor=blue]{hyperref}
\usepackage{cleveref}
\usepackage{comment}
\usepackage{mathtools}
\usepackage{soul}

 \Crefname{equation}{Eq.}{Eqs.}
\Crefname{figure}{Fig.}{Figs.}

\begin{document}
\title{Beyond-mean-field phases of rotating dipolar condensates in the strongly correlated regime}

\author{Paolo Molignini}
\email{paolo.molignini@fysik.su.se}
\affiliation{Department of Physics, Stockholm University, AlbaNova University Center, 10691 Stockholm, Sweden}
\date{\today}

\begin{abstract}
Rotating dipolar Bose-Einstein condensates exhibit rich physics due to the interplay of long-range interactions and rotation, leading to unconventional vortex structures and strongly correlated phases.
While most studies rely on mean-field approaches, these fail to capture quantum correlations that become significant at high rotation speeds and strong interactions. 
In this study, we go beyond the mean-field description by employing a numerically exact multiconfigurational approach to study finite-sized dipolar condensates. 
We reveal novel vortex structures, rotating cluster states, and strong fragmentation effects, demonstrating that beyond-mean-field correlations remain prominent even in larger systems. 
By quantifying deviations from mean-field theory, we provide a predictive framework for analyzing experiments and exploring emergent quantum phases, with implications for both the fundamental theory of ultracold gases and the quantum simulation of correlated superfluid systems like in neutron stars.
\end{abstract}
\maketitle

%%%%%%%%%%%%%%%%%%
%%%%% INTRO %%%%%
%%%%%%%%%%%%%%%%%%
%%%%%%%%%%%%%%%%%%%%%%%%%%%%%%%%%%%%%%%%%%%%%%%%%%%%%%%%%%%%%%%%%%%%%%%%%%%%%%%%%%%%%%%%%%%%%
\section{Introduction} 

The recent progress in experiments with dipolar ultracold atoms~\cite{Griesmaier:2005, Lahaye:2008, Lu:2011, Aikawa:2012} and molecules~\cite{Ni:2008, Molony:2014, Park:2015, Rvachov:2017, Stevenson:2023, Bigagli:2024} has provided a platform for studying novel quantum phases emerging from the interplay between contact and long-range dipole-dipole interactions (DDI)~\cite{Lahaye:2009, Gadway:2016, Moses:2017}.
These systems have enabled the realization of quantum droplets~\cite{Kartashov:2019, Zhang:2021} and supersolid phases~\cite{Lu:2015, Baillie:2018, Tanzi:2019, Chomaz:2019, Guo:2019, Natale:2019, Tanzi:2019-2, Tanzi:2021, Norcia:2021, Sohmen:2021, Biagioni:2022, Recati:2023, Chomaz:2023, Biagioni:2024}, as well as the observation of dipolar quantum gases with metastable properties~\cite{Boettcher:2019, Sanchez-Baena:2023, Zhang:2024}.
These interactions give rise to striking phenomena such as unconventional vortex clustering and dynamics~\cite{Easton:2023, Sabari:2024, Bland:2023, Tengstrand:2019}, roton instabilities~\cite{Hufnagl:2011}, and self-bound quantum droplets~\cite{Zhang:2021, Kartashov:2019}.

Beyond static properties, the interplay between dipolar interactions and external rotation introduces even richer physics.
Recent studies have demonstrated that rotating dipolar gases exhibit complex vortex arrangements, turbulence, and dynamically stabilized cluster states~\cite{Kumar:2016, Bhowmik:2020, Fletcher:2021, Mukherjee:2022, Casotti:2024, Poli:2024, Chatterjee:2024}. 
A particularly intriguing aspect of rotating dipolar systems is their connection to astrophysical phenomena. 
In neutron stars, superfluid vortices exhibit long-range interactions that shape the internal structure and rotational properties of the star~\cite{Poli:2023, Bland:2024}. 
Understanding the interplay between rotation and dipolar interactions is therefore essential for advancing both laboratory-based quantum simulations and our knowledge of astrophysical superfluid systems.

The presence of strong dipolar interactions modifies the well-known vortex lattice structures observed in non-dipolar BECs. 
In contrast to conventional Abrikosov lattices, dipolar condensates can exhibit striped vortex arrangements, fragmented clusters, or even phase-separated vortex distributions~\cite{Bland:2023, Sabari:2024}. 
Theoretical studies have predicted exotic vortex lattice phases, including triangular, square, and amorphous configurations, depending on the interaction strength and rotation frequency~\cite{Mukherjee:2022, Chatterjee:2024}. 
Furthermore, at high rotation speeds, vortex configurations may be torn apart completely.
Under those extreme circumstances, more exotic quantum phases may emerge, as seen in recent experimental and numerical studies identifying quantum Hall-like states~\cite{Fletcher:2021, Mukherjee:2022}.

Despite these advancements, most studies on rotating dipolar condensates have relied on mean-field descriptions based on the extended Gross-Pitaevskii equation (eGPE)~\cite{Brachet:2012}. 
While this approach captures essential features of dipolar Bose-Einstein condensates (BECs), it neglects crucial beyond-mean-field effects that become increasingly important in strongly correlated regimes~\cite{Boettcher:2019}.
In particular, the inclusion of the Lee-Huang-Yang (LHY) correction is often used to partially account for quantum fluctuations. 
The LHY term originates from the first-order correction to the ground-state energy of a weakly interacting Bose gas due to quantum fluctuations, as initially derived by Lee, Huang, and Yang for contact-interacting bosons in the dilute gas limit~\cite{Lee:1957-1,Lee:1957-2}. 
This correction was later extended to dipolar gases, where it introduces an additional density-dependent term in the eGPE, modifying the effective equation of state of the condensate~\cite{Lima:2011, Lima:2012}.

Despite great popularity of the eGPE, the phenomenological inclusion of the LHY correction is not fully self-consistent, as it is derived under the assumption of weak correlations and a homogeneous background density. 
In rapidly rotating condensates, where vortex lattices form and local density variations become significant, the LHY term derived in the homogeneous gas approximation may not accurately describe the system’s behavior \cite{Bisset:2016, Waechtler:2016}. 
Moreover, in regimes where strong interactions lead to the formation of supersolid-like structures or cluster states, the validity of the LHY correction is further compromised, as it does not capture higher-order correlations, effective three-body interactions, or quantum depletion beyond Bogoliubov theory \cite{Chomaz:2019, Hertkorn:2021}.
This limitation underscores the need for beyond-mean-field approaches that can fully account for the many-body correlations in these systems.

In addition to these theoretical limitations, the eGPE may also prove inadequate for describing certain experimental setups.
Recent advances in experimental techniques have made it possible to study ultracold atomic gases consisting of only tens to hundreds of particles. 
Sophisticated trapping methods -- ranging from optical pumping~\cite{Reetz-Lamour:2008-1, Reetz-Lamour:2008-2} and optical tweezers~\cite{Wang:2020} to geometric constrictions~\cite{Brantut:2012} and evaporative cooling~\cite{Zeiher:2021} -- have enabled precise control over small-scale quantum systems.
In these small-scale systems, quantum fluctuations are enhanced due to the reduced particle number and stronger finite-size effects. 
The breakdown of the mean-field description in these cases suggests that fully beyond-mean-field approaches—such as quantum Monte Carlo methods, correlated many-body wavefunctions, or multiconfigurational time-dependent Hartree (MCTDH) methods—are necessary to capture the interplay of rotation, long-range interactions, and strong correlations.

In this work, employing the MCTDH method, we demonstrate that quantum correlations beyond mean-field theory play a crucial role in shaping the ground-state phase diagram of rotating dipolar condensates, even for relatively large particle numbers.
By systematically varying interaction strength and rotation frequency for particle numbers from $N=5$ to $N=50$, we uncover a plethora of novel vortex structures, extended droplets, rotating cluster phases, and correlation-driven effects that cannot be captured by mean-field approximations. 
To quantify the deviations from mean-field theory, we compute an effective orbital occupation order parameter.
Remarkably, we find that even in systems with a larger particle number beyond-mean-field effects become prominent when rotations are fast and/or interactions are strong.

Our findings provide crucial insights into the stability and dynamics of dipolar condensates under rotation, offering a quantitative framework for analyzing experiments with finite-sized ensembles of bosons. This, in turn, paves the way for future quantum simulations of strongly correlated dipolar gases, with implications ranging from condensed matter systems to astrophysical superfluidity.

The rest of this paper is structured as follows.
In section \ref{sec:system}, we describe the setup we study.
In section \ref{sec:methods}, we give a brief overview of the computational method and the observables we employ to investigate the system numerically exactly.
In sections \ref{sec:results}, we present our results for a varying particle number between $N=5$ and $N=50$ particles, in terms of ground-state energies, density distributions, and fragmentation, and discuss their dependence on the degree of beyond-mean-field corrections.
Section \ref{sec:conclusions} summarizes our study and provides an outlook on potential future research directions.
%%%%%%%%%%%%%%%%%%%%%%%%%%%%%%%%%%%%%%%%%%%%%%%%%%%%%%%%%%%%%%%%%%%%%%%%%%%%%%%%%%%%%%%%%%%%%

%%%%%%%%%%%%%%%%%%
%%%%% SYSTEM %%%%%
%%%%%%%%%%%%%%%%%%
%%%%%%%%%%%%%%%%%%%%%%%%%%%%%%%%%%%%%%%%%%%%%%%%%%%%%%%%%%%%%%%%%%%%%%%%%%%%%%%%%%%%%%%%%%%%%
\section{System and protocol}
\label{sec:system}
We set out to investigate the ground-state properties of \( N \) dipolar-interacting bosons of mass \( m \) in the presence of a rotating potential.  
The bosons are confined in a two-dimensional (2D) geometry described by the coordinates $\mathbf{r}_j=(x_j,y_j)$ for each particle.
The total Hamiltonian of this system can be written in the rotating frame as  
\begin{equation}
\mathcal{H} = \sum_{j=1}^N \left[ T(\mathbf{r}_j) + V(\mathbf{r}_j) \right] + \sum_{j<k}^N W(\mathbf{r}_j -\mathbf{r}_k).
\label{eq:Ham1}
\end{equation}
In the equation above,
\begin{equation}
T(\mathbf{r}) = -\frac{\hbar^2}{2 m}\nabla_{\mathbf{r}}^2 - \Omega L_z
\end{equation}
is the kinetic energy that includes a rotational contribution of angular speed $\Omega$ from the angular momentum $L_z = x p_y - y p_x$, which, in position representation, takes the form  
\begin{equation}
L_z = -i\hbar \left( x \frac{\partial}{\partial y} - y \frac{\partial}{\partial x} \right).
\end{equation}
The second term in the one-body part of the Hamiltonian is the trapping potential, which we take to be an isotropic harmonic trap of the form
\begin{equation}
    V(\mathbf{r}) = \frac{1}{2} \omega \mathbf{r}^2.
\end{equation}
The particles additionally feel a regularized dipole-dipole repulsion of strength $g_d$, effectively described by 
\begin{equation}
    W(\mathbf{r}_j - \mathbf{r}_k) = \frac{g_d}{|\mathbf{r}_j - \mathbf{r}_k|^3 + \alpha},
\end{equation}
We note that -- while a self-consistent dimensional reduction of the dipole-dipole interactions would lead to a normalization with Bessel functions~\cite{Sinha:2007, Ticknor:2009, Cremon:2010, Cai:2010, Deuretzbacher:2010, Yousefi:2015} -- the simple regularization parameter $\alpha=0.05$ captures the short-range cutoff induced by transverse confinement equally well.
Moreover, it ensures numerical stability without altering the qualitative phase behavior due to the universality of the dipole-dipole interactions~\cite{Hofmann:2021}.
The regularization parameter induces a finite interaction in the limit $|\mathbf{r}_j - \mathbf{r}_k| \to 0$ and thus gives rise to a soft-core behavior that -- in static settings -- is known to induce density-modulated ground states~\cite{Cinti:2010, Cinti:2014, Biagioni:2018-thesis, Blakie:2024, Poli:2024-PRA}.
In experiments, the interactions can be tuned via various mechanisms including confinement induced resonances~\cite{Tiesinga:1992, Tiesinga:2000, Bergeman:2003,Saeidian:2008} and magnetic or electric field manipulation~\cite{Tang:2018, Li:2021, Giovanazzi:2022}.
In particular, isotropic two-dimensional interactions can be achieved when the dipoles are confined to a plane and are all aligned perpendicularly to it.
In our numerics, we adopt natural units corresponding to $\hbar=m=1$, and scale the Hamiltonian \eqref{eq:Ham1} by $\frac{\hbar^2}{mL^2}$, where $L$ is the natural length scale of the harmonic trap.

%%%%%%%%%%%%%%%%%%%%%%%%%%%%%%%%%%%%%%%%%%%%%%%%%%%%%%%%%%%%%%%%%%%%%%%%%%%%%%%%%%%%%%%%%%%%%

%%%%%%%%%%%%%%%%%%%
%%%%% METHODS %%%%%
%%%%%%%%%%%%%%%%%%%
%%%%%%%%%%%%%%%%%%%%%%%%%%%%%%%%%%%%%%%%%%%%%%%%%%%%%%%%%%%%%%%%%%%%%%%%%%%%%%%%%%%%%%%%%%%%%
\section{Methods}
\label{sec:methods}
To extract information from the rotating-frame Hamiltonian Eq.~\eqref{eq:Ham}, we solve the corresponding Schr\"{o}dinger equation
\begin{equation}
    i \hbar \partial_t \Psi(\mathbf{r}, t) = \mathcal{H} \Psi(\mathbf{r}, t)
\end{equation}
with  the MultiConfigurational Time-Dependent Hartree method for indistinguishable particles (MCTDH)~\cite{Streltsov:2006, Streltsov:2007, Alon:2007, Alon:2008, Lode:2012, Lode:2016, Fasshauer:2016}.
The MCTDH approach assembles the many-body wave function $\Psi(\mathbf{r}, t)$ as a dynamically optimized superposition of permanents.
The permanents are in turn constructed from $M$ variationally determined single-particle functions, known as \emph{orbitals}.
The method self-consistently refines both the expansion coefficients and the orbital basis, facilitating ground-state computations via imaginary time relaxation or real-time evolution for dynamical studies.
In this work, we focus exclusively on imaginary time propagation to ensure that the system reaches its lowest energy state.
For numerical computations, we employ the MCTDH-X software suite~\cite{Lin:2020,Lode:2020,Molignini:2025-SciPost,MCTDHX}, which represents the state-of-the-art implementation of the MCTDH formalism.
A detailed exposition of the method can be found in Appendix~\ref{app:MCTDHX}.

By employing a many-body ansatz with multiple orbitals, we are able to capture fragmented states such as cluster states, which remain inaccessible to mean-field theories and their extensions, including the eGPE.
While exact convergence is achieved in the limit $M \to \infty$, practical calculations require only a finite number of orbitals to obtain numerically exact results -- i.e., solutions that remain unchanged upon further increasing $M$.
In this study, we are interested in the breakdown of the mean-field calculations and will therefore perform simulations for increasing orbital number up to $M=10$.

To describe the system’s ground state in the rotating frame and measure the deviation from mean-field results, we extract a collection of quantities from the variational many-body wave function $\left| \Psi \right>$.
The simplest quantity to track is the total energy
\begin{align} E = \left< \Psi \right| \hat{H} \left| \Psi \right>, 
\end{align} 
which provides insight into the overall ground-state convergence of our simulations as the number of orbitals is increased.
The time-dependent variational principle ensures that, within a given variational class, lower energy corresponds to a more accurate approximation of the exact ground state.

Another crucial measure of many-body convergence is given by the orbital occupation, i.e. how effectively the particles are using the provided variational subspace. 
This quantity can be computed as follows.
First, we calculate the reduced one-body density matrix (1-RDM), defined as 
\begin{align} \rho^{(1)}(\mathbf{r},\mathbf{r}') &= \left< \Psi \right| \hat{\Psi}^{\dagger}(\mathbf{r}) \hat{\Psi}(\mathbf{r}') \left| \Psi \right>. 
\end{align} 
From the 1-RDM, we can obtain information about the orbital occupation by performing a spectral decomposition
\begin{equation} 
\rho^{(1)}(\mathbf{r},\mathbf{r}') = \sum_i \rho_i \phi^{(\mathrm{NO}),*}_i(\mathbf{r}')\phi^{(\mathrm{NO})}_i(\mathbf{r}). 
\label{eq:RDM1} 
\end{equation} 
In this equation, $\rho_i$ denote the orbital occupations, which are the eigenvalues of the 1-RDM.
By convention, the occupations $\rho_i$ are ordered from largest to smallest.
They quantify the population of the natural orbitals $\phi_i^{(\mathrm{NO})}$, i.e. the eigenfunctions of the 1-RDM.

The occupations already provide insight into the degree of wave function fragmentation in the system.
For example, if only a single orbital is macroscopically occupied ($\rho_1 \approx 1$), then the wave function is well-described by a coherent  mean-field solution.
If $\rho_1 < 1$, fragmentation into multiple orbitals occurs.
To quantify this fragmentation more precisely, we introduce the inverse participation ratio,
\begin{equation}
    \mathrm{IPR} = \frac{1}{\sum_{j=1}^M \rho_j^2}.
\end{equation}
The IPR better quantifies how many orbitals contribute to the total wave function.
For example, a mean-field solution with $\rho_1=1$ and $\rho_{>1}=0$ leads to $\mathrm{IPR}=1$.
At the other extreme, if the ground state is equally split between all $M$ orbitals, i.e. $\rho_1 = \rho_2 = \cdots = \rho_M = \frac{1}{M}$, we have $\mathrm{IPR} = \frac{1}{\frac{1}{M^2} M} = M$ and the entire variational subspace is filled.
Therefore, tracking the value of $\mathrm{IPR}$ against the total available orbital number $M$ provides a clear indication of how many orbitals are, on average, needed to describe the many-body state. 

In highly correlated regimes, calculations with an increasing number of orbitals $M$ can lead to different states until convergence in $M$ sets in.
To distinguish these different states visually, we directly calculate the one-body density distribution of the bosonic system, 
\begin{align} \rho(\mathbf{r}) &= \left< \Psi \right| \hat{\Psi}^{\dagger}(\mathbf{r}) \hat{\Psi}(\mathbf{r}) \left| \Psi \right>, 
\end{align} 
where $\Psi^{(\dagger)}(\mathbf{r})$ denotes the bosonic creation (annihilation) operator at position $\mathbf{r}$.
Based on the spatial distribution of the particles and the orbital occupation, we will construct phase diagrams in the parameter space spanned by the interaction strength $g_d$ and the rotation speed $\Omega$.

%%%%%%%%%%%%%%%%%%%%%%%%%%%%%%%%%%%%%%%%%%%%%%%%%%%%%%%%%%%%%%%%%%%%%%%%%%%%%%%%%%%%%%%%%%%%%

%%%%%%%%%%%%%%%%%%%
%%%%% RESULTS %%%%%
%%%%%%%%%%%%%%%%%%%
%%%%%%%%%%%%%%%%%%%%%%%%%%%%%%%%%%%%%%%%%%%%%%%%%%%%%%%%%%%%%%%%%%%%%%%%%%%%%%%%%%%%%%%%%%%%%
\section{Results}
\label{sec:results}

We now present the results of our many-body simulations for different particle numbers and orbitals.  
We start with the \( N=50 \) case—which is expected to behave more like mean-field theory due to the larger particle number—and progressively reduce the particle number to \( N=20 \), \( N=10 \), and finally \( N=5 \) to study systems where quantum fluctuations are more predominant.  
For each case, we discuss our results in terms of ground-state energy, inverse participation ratio (IPR), and density distributions.

Overall, we find at least eight qualitatively different states in the density distribution, which we briefly summarize here before presenting our detailed results.  
These different states are highlighted in different colors in our phase diagrams.

\emph{Superfluid} states (marked in white) are characterized by a condensed atomic cloud in a Thomas-Fermi shape~\cite{Dalfovo:1999} and an occupation of a single orbital near unity.

When more orbitals become occupied, the state transitions to a \emph{fragmented superfluid}.  
In this state, the overall Thomas-Fermi spatial profile may survive or may be partially deformed.  
We do not distinguish between these two cases and mark both in light blue in our phase diagrams.

As the rotation becomes stronger, it might become energetically favorable for the superfluid cloud to embed vortices in its core.  
Vortices are characterized by a phase winding of the superfluid order parameter around the vortex core and manifest in the density as depletion at the vortex center~\cite{Fetter:2001}.  
For sufficiently large numbers of vortices, they arrange into a triangular \emph{Abrikosov lattice}~\cite{Abrikosov:1957, Baym:2004}, which may be distorted due to finite-size effects and interactions~\cite{Komineas:2007, Fetter:2008, Fetter:2009}.  
In this work -- to more precisely map the onset of vortex physics -- we distinguish between \emph{single-vortex} occurrences (marked in cyan in the phase diagrams) and \emph{multiple-vortex} states (marked in green).

Vortices can coexist with other types of inhomogeneities in the density.  
We find that the most common manifestation of such inhomogeneities is the accumulation of density \emph{clusters} in the center of the rotating trap.  
The regions in parameter space where these states appear are marked in yellow in our phase diagram.

However, we find that clusters do not always coexist with vortices.  
This is particularly true for systems with a lower number of particles.  
Sometimes, clusters appear surrounded by a delocalized ``aura'' of particles, which we highlight in orange in our phase diagrams.  
Some other times -- typically at larger interactions or rotation frequencies -- the clusters appear as standalone structures.  
These cases are denoted in red.

Finally, it may occur that rotation frequencies close to or at the critical threshold \( \Omega = \omega \) disrupt the atomic cloud completely and lead to the particles escaping the traps.  
This results in a chaotic density pattern which we do not consider as converged in our numerics.
We mark these cases in black in the phase diagrams.

\subsection{$N=50$ bosons}
\label{sec:N=50}

%%%%%%%%%%%%%%%%%%%%%%%%%%%%%%%%%%%%%%%%%%%%
\begin{figure}[t!]
\centering
\includegraphics[width=1.0\columnwidth]{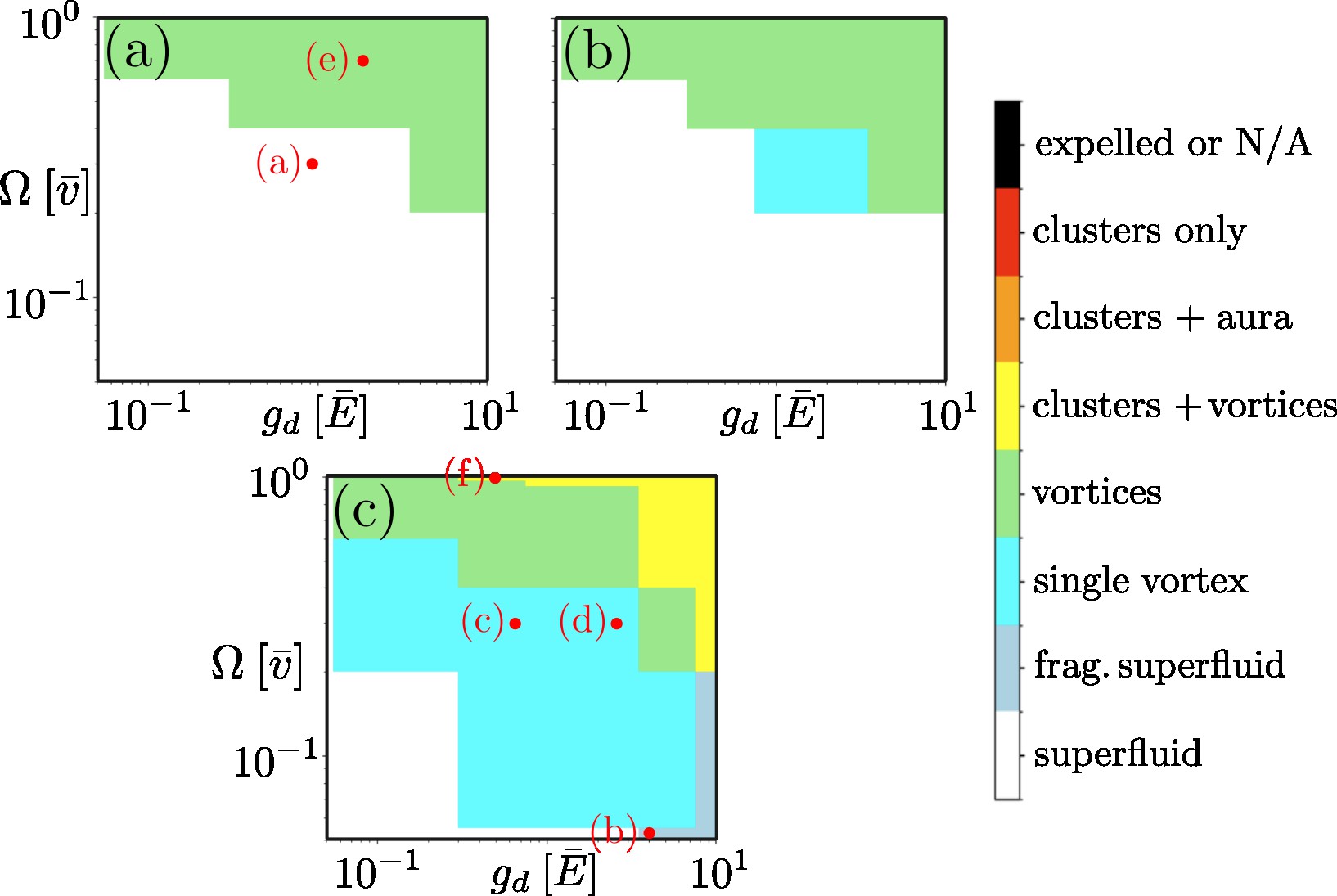}
\caption{
The different phases obtained from the simulation of $N=50$ dipolar-interacting bosons in a rotating potential as a function of interaction strength $g_d$ and rotation speed $\Omega$ (on a logarithmic scale).
The different panels show results obtained for different number of orbitals:
(a) $M=1$ orbitals,
(b) $M=2$ orbitals,
(c) $M=5$ orbitals.
The red letters in the phase diagram refer to the points in parameter space where the density distributions of Fig.~\ref{fig:density-N-50} were probed.
}
\label{fig:PD-N-50}
\end{figure}
%%%%%%%%%%%%%%%%%%%%%%%%%%%%%%%%%%%%%%%%%%%%

\begin{figure}[t!]
\centering
\includegraphics[width=0.9\columnwidth]{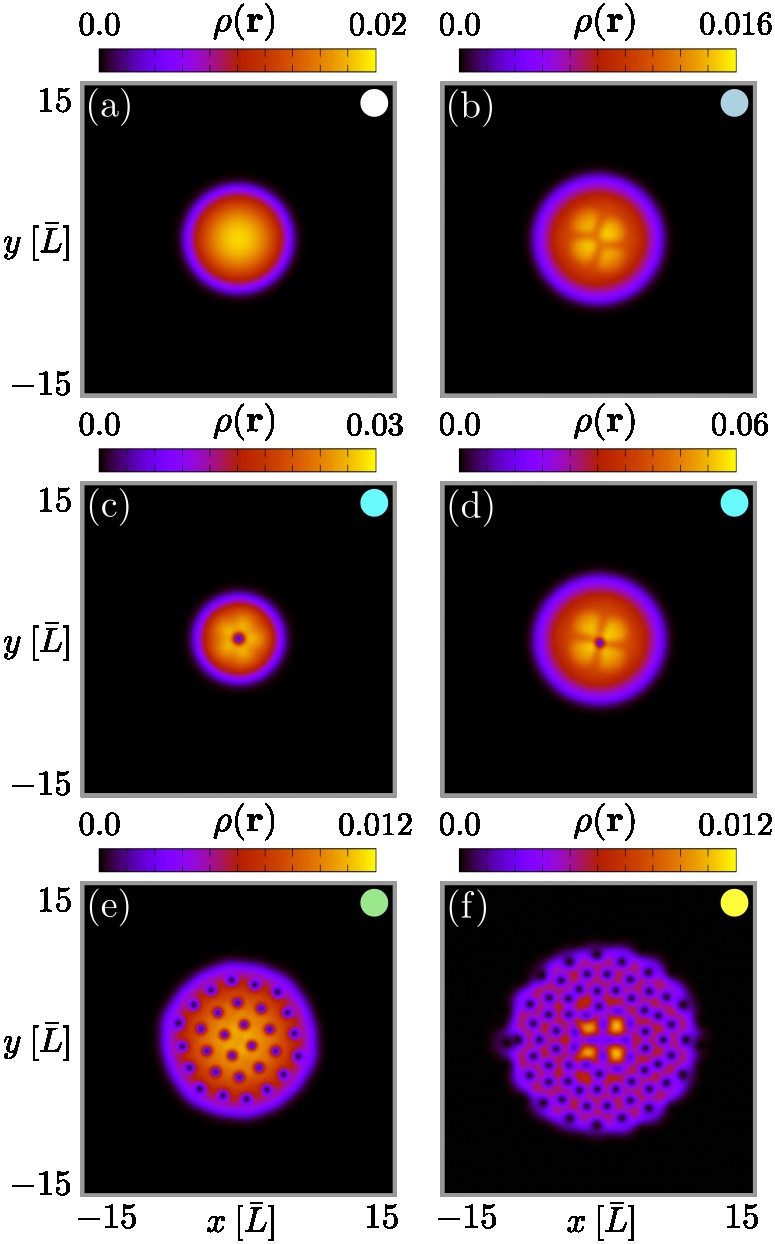}
\caption{
The different states appearing for $N=50$ dipolar bosons in a rotating trap:
(a) superfluid,
(b) fragmented superfluid with angular roton profile,
(c) superfluid with a single vortex core,
(d) fragmented superfluid with a single vortex core,
(e) Abrikosov lattice of vortices,
(f) coexistence of deformed vortex lattice with central clusters.
The circles in the upper-right corners have the same color code used in the phase diagram ~\ref{fig:PD-N-50} and indicate which states the density represents. 
}
\label{fig:density-N-50}
\end{figure}

Fig.~\ref{fig:PD-N-50} visualizes the kind of states that appear for $N=50$ dipolar-interacting bosons as a function of interaction strength $g_d$ and rotation speed $\Omega$ (note the logarithmic scale).
The different panels refer to calculations performed for increasing number of orbitals: (a) $M=1$ -- equivalent to a GPE, (b) $M=2$ and (c) $M=5$, which is a soft limit in the number of orbitals that can be employed with current computational infrastructures.
The phase diagrams illustrate that mean-field approximations and small deviations from it ($M=2$) predict a large portion of the parameter space at slow-enough rotations ($\Omega \lesssim 0.5$) being occupied by pure superfluid states.
Fig.~\ref{fig:density-N-50}(a) depicts an example of the density distribution for these superfluid states exhibiting the typical Thomas-Fermi profile.
At larger rotation speeds, vortices start to appear, with stronger interactions favoring vortices at slower speeds as low as $\Omega = 0.3 \bar{v}$ for $g_d = 10 \bar{E}$.
Typically, the vortices arrange themselves into a regular Abrikosov lattice, as shown in Fig.~\ref{fig:density-N-50}(e).

%%%%%%%%%%%%%%%%%%%%%%%%%%%%%%%%%%%%%%%%%%%%
%%%%%%%%%%%%%%%%%%%%%%%%%%%%%%%%%%%%%%%%%%%%

At slow rotations and weak interactions, accounting for more quantum fluctuations by using $M=5$ orbitals leads to a similar phase diagram, as shown in Fig.~\ref{fig:PD-N-50}(c).
A notable exception is that the pure superfluid phase is actually restricted to a much narrower region in parameter space.
Much of the parameter space is instead occupied by states with single vortices -- with a density profile depicted in Fig.~\ref{fig:density-N-50}(c).

On top of this quantitative shift, though, using a larger number of orbitals also reveals \emph{qualitative} differences in regimes of strong interactions, depicted in light blue and yellow in Fig.~\ref{fig:PD-N-50}(c).
The shape of these strongly interacting states depend on the rotational frequency.
At slow rotations, the states appear as fragmented superfluids.
These states display four-fold symmetric anisotropies in the density characteristic of angular roton softening~\cite{Schmidt:2021}, as depicted in Fig.~\ref{fig:density-N-50}(b), and can also coexist with single vortices in the center of the condensate, as shown in  Fig.~\ref{fig:density-N-50}(d).
The roton pattern appearing in the density is a precursor of a supersolid ordering that occurs at faster rotations~\cite{Lahrz:2014, Alana:2021, Schmidt:2021}.
In fact, in the strong interacting and fast rotating regime, we do observe the coexistence of clusters -- arranged in a square lattice -- with an underlying disordered vortex lattice.
A prototypical configuration for this state is pictured in Fig.~\ref{fig:density-N-50}(f).

%%%%%%%%%%%%%%%%%%%%%%%%%%%%%%%%%%%%%%%%%%%%
\begin{figure}[t!]
\centering
\includegraphics[width=0.8\columnwidth]{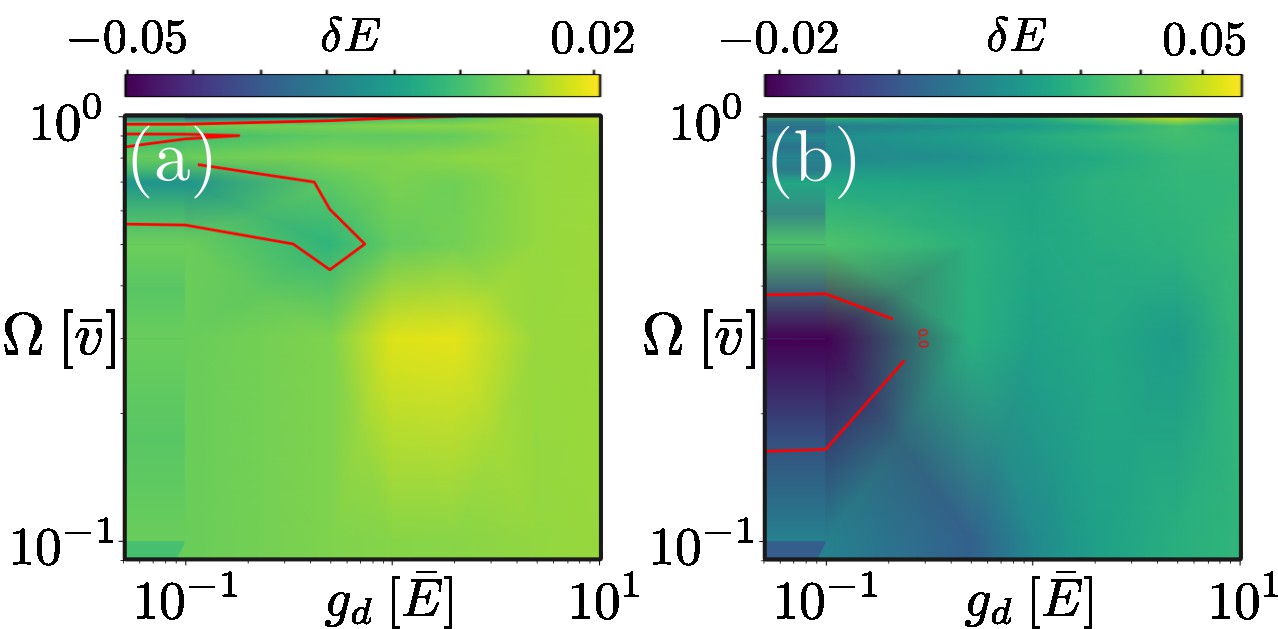}
\caption{
The difference in ground state energy between different $N=50$ calculations with increasing orbital number: (a) $\delta E = \frac{E_{M=1} - E_{M=2}}{E_{M=1}}$, (b) $\delta E = \frac{E_{M=2} - E_{M=5}}{E_{M=2}}$.
The red lines indicate the $\delta E = 0$ contours.
}
\label{fig:energies-N-50}
\end{figure}
%%%%%%%%%%%%%%%%%%%%%%%%%%%%%%%%%%%%%%%%%%%%
%%%%%%%%%%%%%%%%%%%%%%%%%%%%%%%%%%%%%%%%%%%%
\begin{figure}[t!]
\centering
\includegraphics[width=0.8\columnwidth]{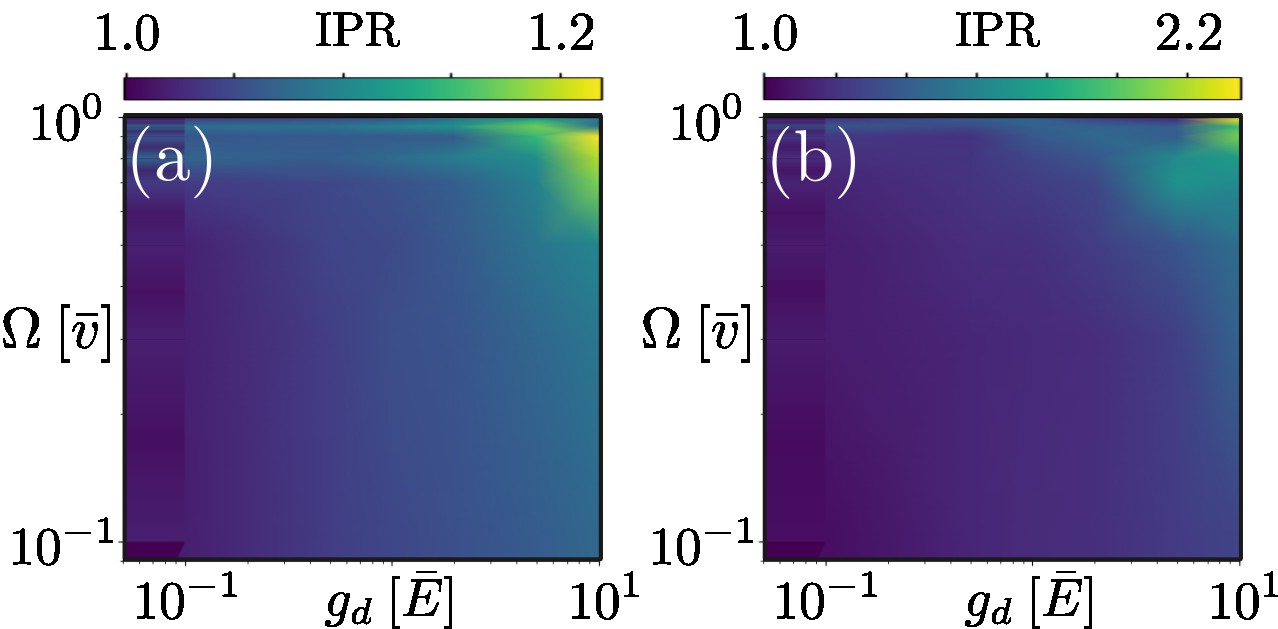}
\caption{
The IPR in parameter space for $N=50$ calculations with (a) $M=2$ and (b) $M=5$ orbitals.
}
\label{fig:ipr-N-50}
\end{figure}
%%%%%%%%%%%%%%%%%%%%%%%%%%%%%%%%%%%%%%%%%%%%

A legitimate question to ask is whether our numerical calculations with a larger number of orbitals do lead to more precise ground states.
We answer this questions affirmatively by directly comparing the ground-state energy between $M=2$ and $M=1$ calculations in Fig.~\ref{fig:energies-N-50}(a) and between $M=5$ and $M=2$ in Fig.~\ref{fig:energies-N-50}(b).
With the exception of a few regions at weaker interactions (delineated by the red contour lines), we find that increasing the orbital number systematically leads to ground states with lower energy, i.e. we observe energy convergence.
In particular, in regimes of strong interactions $g_d \gtrsim 5 \bar{E}$, including more quantum fluctuations can lead to ground states that are noticeably lower in energy by up to 5\%.
These results indicate that the true ground states of strongly-interacting dipolar bosons under rotations are fundamentally different and energetically gapped from the states accessible with mean-field approximations.
These ground states can be systematically accessed with optimized variational methods such as MCTDH.

To quantitatively assess the departure from the mean-field approximation, Fig.~\ref{fig:ipr-N-50} presents the inverse participation ratio (IPR) for (a) calculations with $M=2$ and (b) calculations with $M=5$. 
The IPR provides a measure of the degree of correlation by quantifying the effective number of orbitals participating in the variational representation the many-body state.
For $M=2$, the IPR remains close to the mean-field value of $\mathrm{IPR} = 1$, indicating only minor deviations. 
However, increasing the variational subspace to $M=5$ orbitals significantly impacts the strongly interacting and fast-rotating regimes, where the many-body state requires, on average, more than two orbitals for an accurate description.
These regions in parameter space are precisely the ones that give rise to a coexistence of vortices and clusters.
These findings suggest that accounting for correlations beyond mean-field approximations is crucial to probe the correct physics of fast-rotating and strongly-interacting dipolar gases.

%%%%%%%%%%%%%%%%%%%%%%%%%%%%%%%%%%%%%%%%%%%%
\begin{figure}[t!]
\centering
\includegraphics[width=1.0\columnwidth]{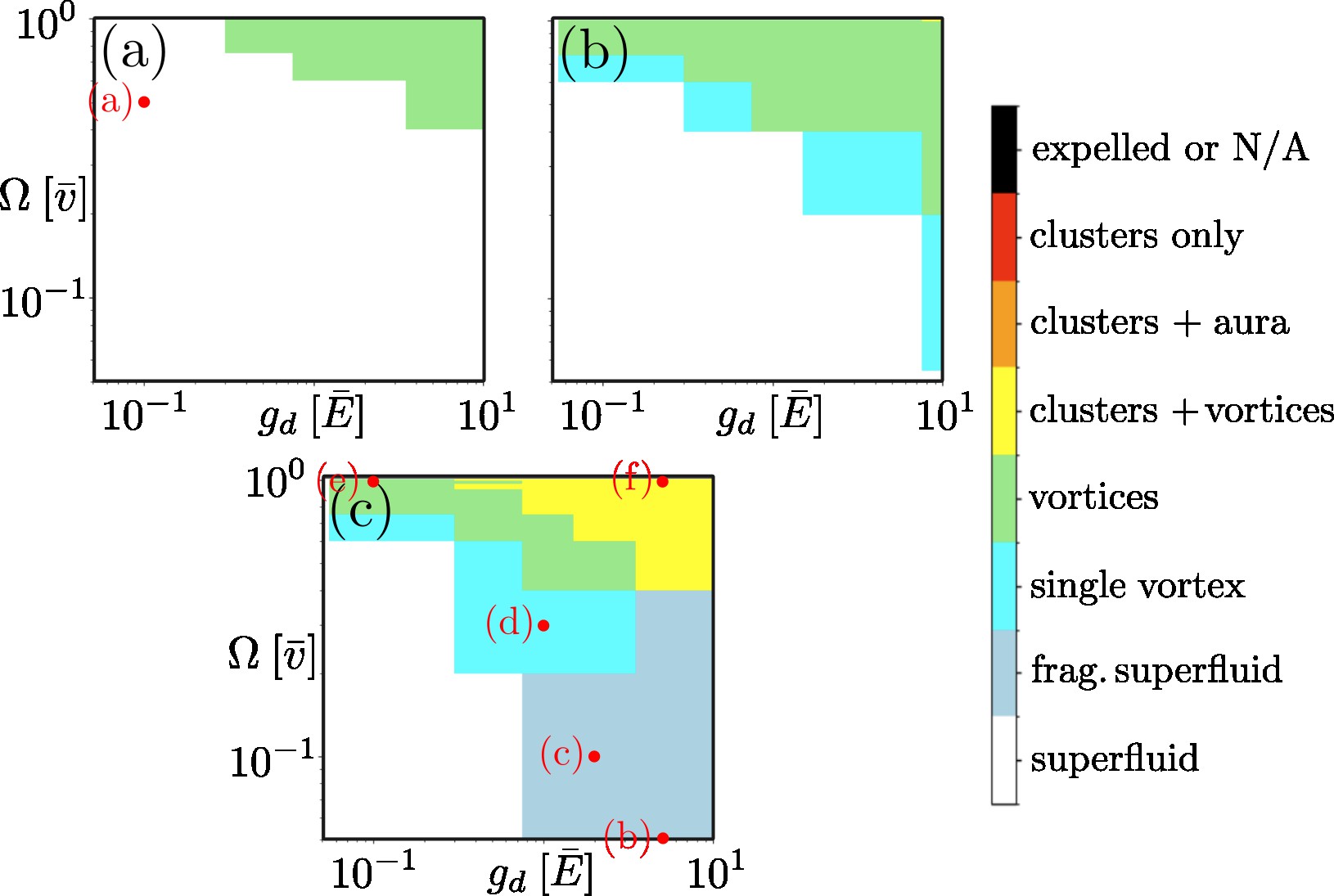}
\caption{
The different phases obtained from the simulation of $N=20$ dipolar-interacting bosons in a rotating potential as a function of interaction strength $g_d$ and rotation speed $\Omega$ (on a logarithmic scale).
The different panels show results obtained for different number of orbitals:
(a) $M=1$ orbitals,
(b) $M=2$ orbitals,
(c) $M=5$ orbitals.
The red letters in the phase diagram refer to the points in parameter space where the density distributions of Fig.~\ref{fig:density-N-20} were probed.
}
\label{fig:PD-N-20}
\end{figure}
%%%%%%%%%%%%%%%%%%%%%%%%%%%%%%%%%%%%%%%%%%%%

\subsection{$N=20$ bosons}
\label{sec:N=20}

%%%%%%%%%%%%%%%%%%%%%%%%%%%%%%%%%%%%%%%%%%%%
\begin{figure}[t!]
\centering
\includegraphics[width=0.9\columnwidth]{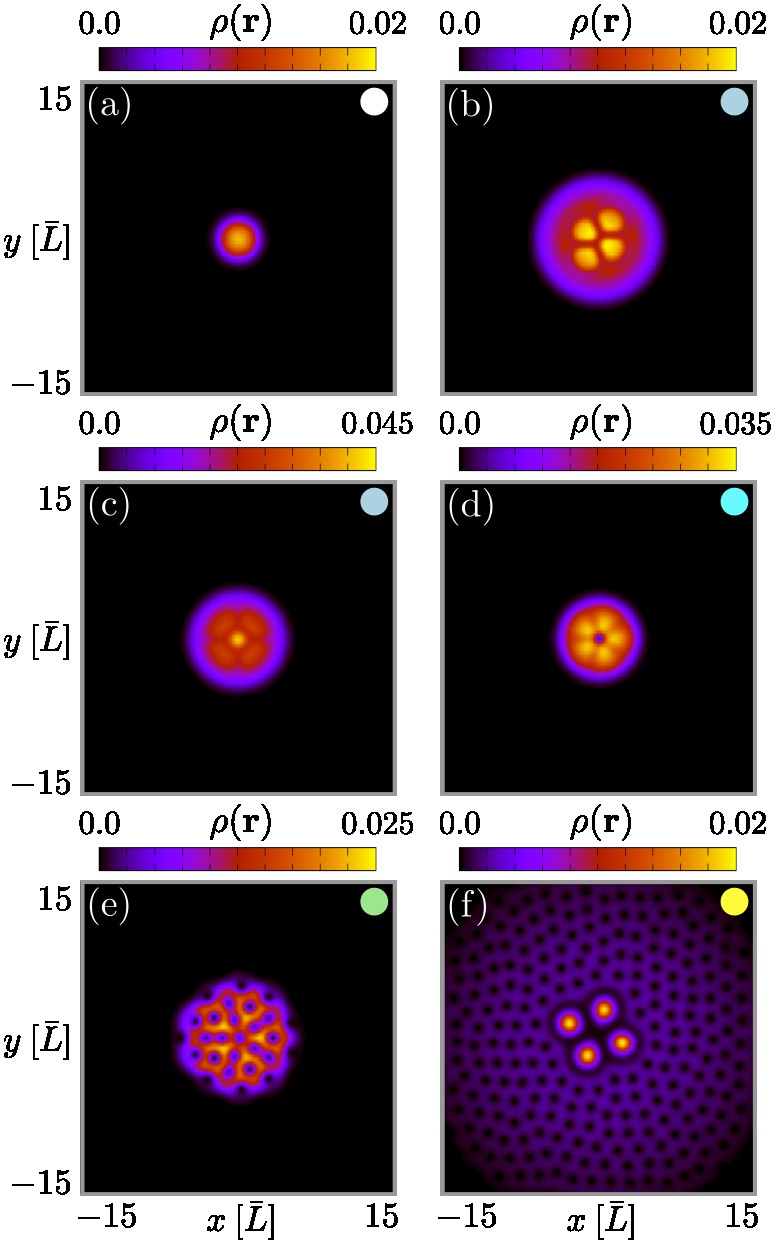}
\caption{
The different states appearing for $N=20$ dipolar bosons in a rotating trap:
(a) superfluid,
(b) fragmented superfluid with angular roton profile,
(c) fragmented superfluid with radial roton profile,
(d) fragmented superfluid with a single vortex core,
(e) disordered vortex lattice,
(f) coexistence of vortex lattice with central clusters.
The circles in the upper-right corners have the same color code used in the phase diagram ~\ref{fig:PD-N-20} and indicate which states the density represents. 
}
\label{fig:density-N-20}
\end{figure}
%%%%%%%%%%%%%%%%%%%%%%%%%%%%%%%%%%%%%%%%%%%%

We now consider systems with a slightly smaller number of particles, $N=20$, to increase the deviation from mean-field regimes.
Fig.~\ref{fig:PD-N-20} illustrates the corresponding phase diagrams for (a) $M=1$, (b) $M=2$, and (c) $M=5$.
The overall behavior is quite similar to the $N=50$ results, but the fragmented superfluid region and the region hosting coexisting clusters and vortices [Fig.~\ref{fig:PD-N-20}(c)] is already much larger.
This indicates a stronger deviation from the states that can be described by mean-field approximations in a larger chunk of the parameter space.

In fact, the density profiles depicted in Fig.~\ref{fig:density-N-20} are fairly similar to the ones obtained for the $N=50$ calculations, with a few notable exceptions.
The roton profile of the fragmented superfluids is stronger [\ref{fig:density-N-20}(b)], other fragmented states with a profile reminiscent of a radial roton~\cite{Schmidt:2021} appear [\ref{fig:density-N-20}(c)], and the vortex lattices become more disordered [\ref{fig:density-N-20}(e)].

\begin{figure}[t!]
\centering
\includegraphics[width=0.8\columnwidth]{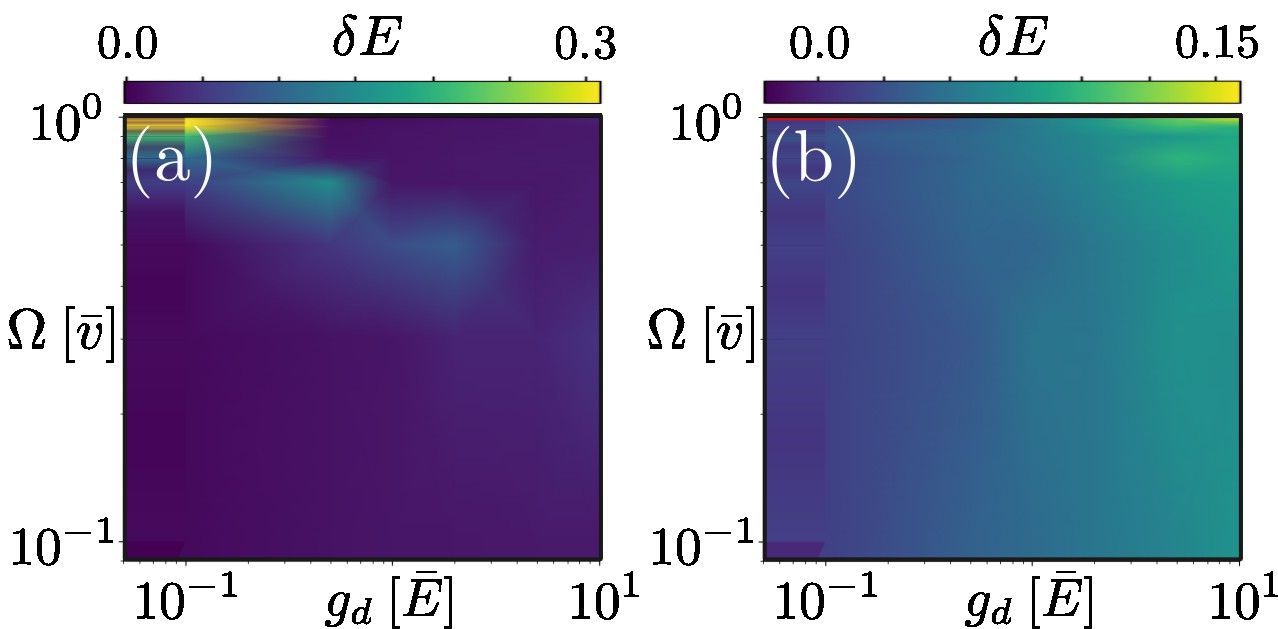}
\caption{
The difference in ground state energy between different $N=20$ calculations with increasing orbital number: (a) $\Delta E = \frac{E_{M=1} - E_{M=2}}{E_{M=1}}$, (b) $\Delta E = \frac{E_{M=2} - E_{M=5}}{E_{M=2}}$.
}
\label{fig:energies-N-20}
\end{figure}

As for the $N=50$ calculations, we find that the energies of the multi-orbital calculations are systematically lower than the mean-field ones.
This is reported in Fig.~\ref{fig:energies-N-20}(a)-(b), where the maximal energy reduction is extremely large and amounts to 15\% in the fast-rotating regime.

%%%%%%%%%%%%%%%%%%%%%%%%%%%%%%%%%%%%%%%%%%%%
%%%%%%%%%%%%%%%%%%%%%%%%%%%%%%%%%%%%%%%%%%%%

\begin{figure}[t!]
\centering
\includegraphics[width=0.8\columnwidth]{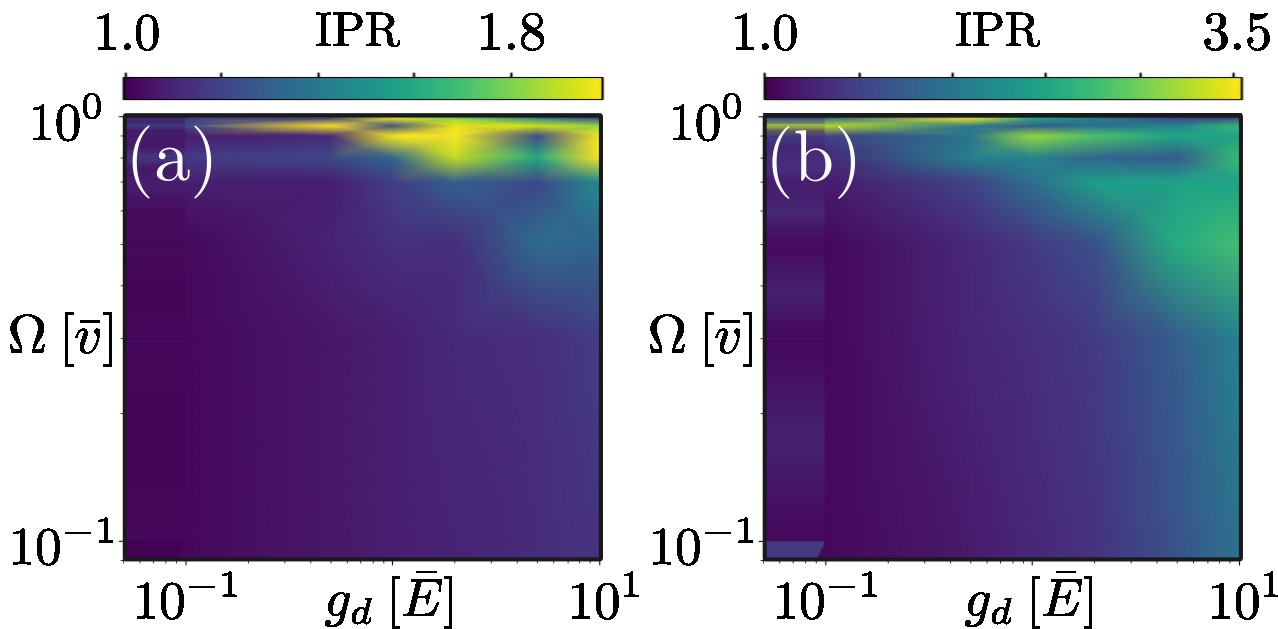}
\caption{
The IPR in parameter space for $N=20$ calculations with (a) $M=2$ and (b) $M=5$ orbitals.
}
\label{fig:ipr-N-20}
\end{figure}

Furthermore, we find from the IPR plotted in Fig.~\ref{fig:ipr-N-20} that the average orbital occupation in this case is much larger, with peaks of around 2 full orbitals for $M=2$ (thereby saturating the entire variational subspace) and 3.5 orbitals for $M=5$.
In particular, the saturation of the $M=2$ subspace indicates that a much larger number of orbitals is required to correctly describe the physics of the fast-rotating and strongly-interacting regimes.

%%%%%%%%%%%%%%%%%%%%%%%%%%%%%%%%%%%%%%%%%%%%
%%%%%%%%%%%%%%%%%%%%%%%%%%%%%%%%%%%%%%%%%%%%

\subsection{$N=10$ bosons}
\label{sec:N=10}

%%%%%%%%%%%%%%%%%%%%%%%%%%%%%%%%%%%%%%%%%%%%
\begin{figure}[t!]
\centering
\includegraphics[width=1.0\columnwidth]{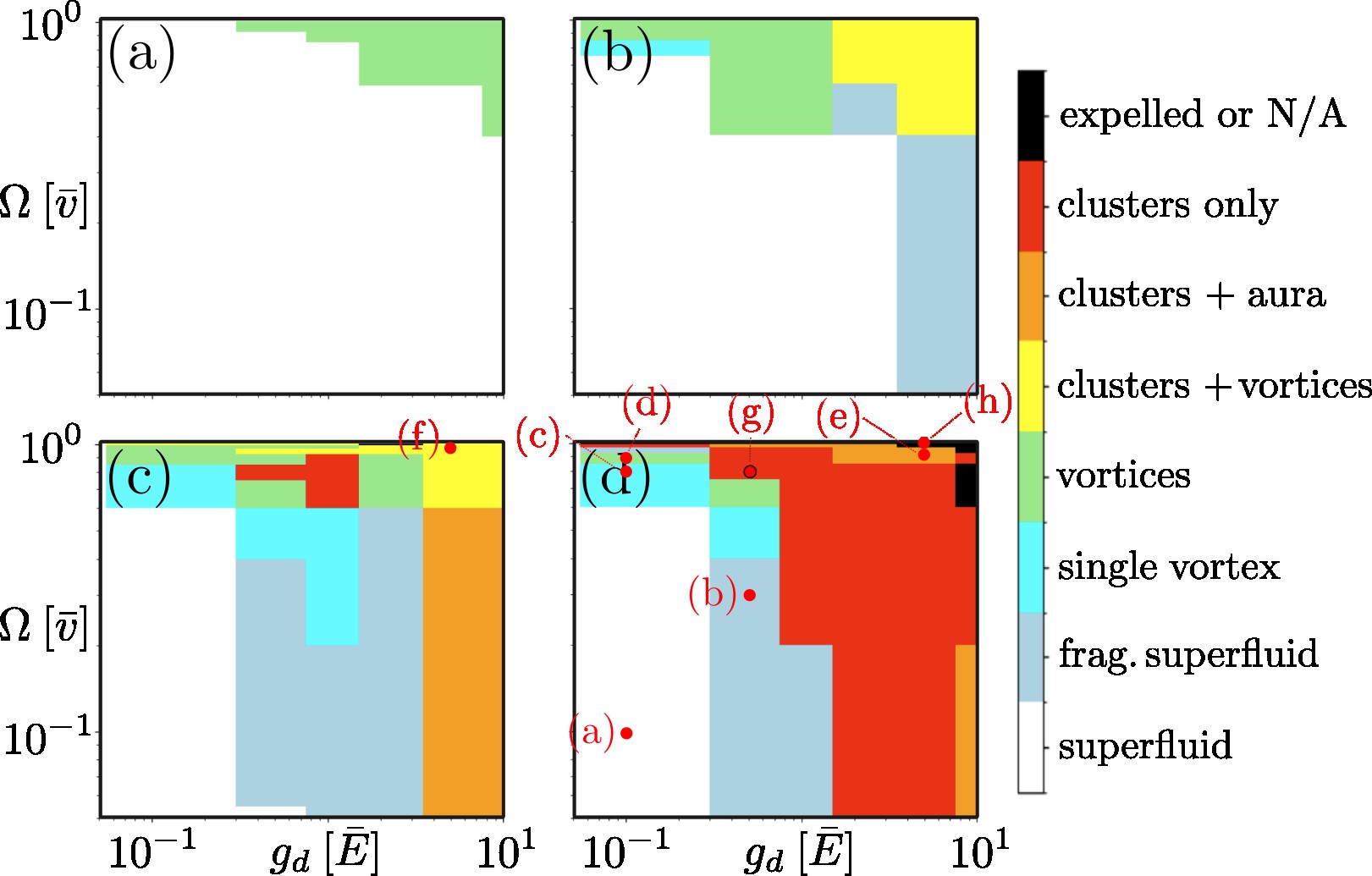}
\caption{
The different phases obtained from the simulation of $N=10$ dipolar-interacting bosons in a rotating potential as a function of interaction strength $g_d$ and rotation speed $\Omega$ (on a logarithmic scale).
The different panels show results obtained for different number of orbitals:
(a) $M=1$ orbitals,
(b) $M=2$ orbitals,
(c) $M=5$ orbitals.
(d) $M=10$ orbitals.
The red letters in the phase diagram refer to the points in parameter space where the density distributions of Fig.~\ref{fig:density-N-10} were probed.
}
\label{fig:PD-N-10}
\end{figure}
%%%%%%%%%%%%%%%%%%%%%%%%%%%%%%%%%%%%%%%%%%%%

We fully enter a non-mean-field regime when we further lower the particle number to $N=10$.
The lower particle number simultaneously enables us to include a higher number of orbitals up to $M=10$ in our calculations.

\begin{figure}[t!]
\centering
\includegraphics[width=0.9\columnwidth]{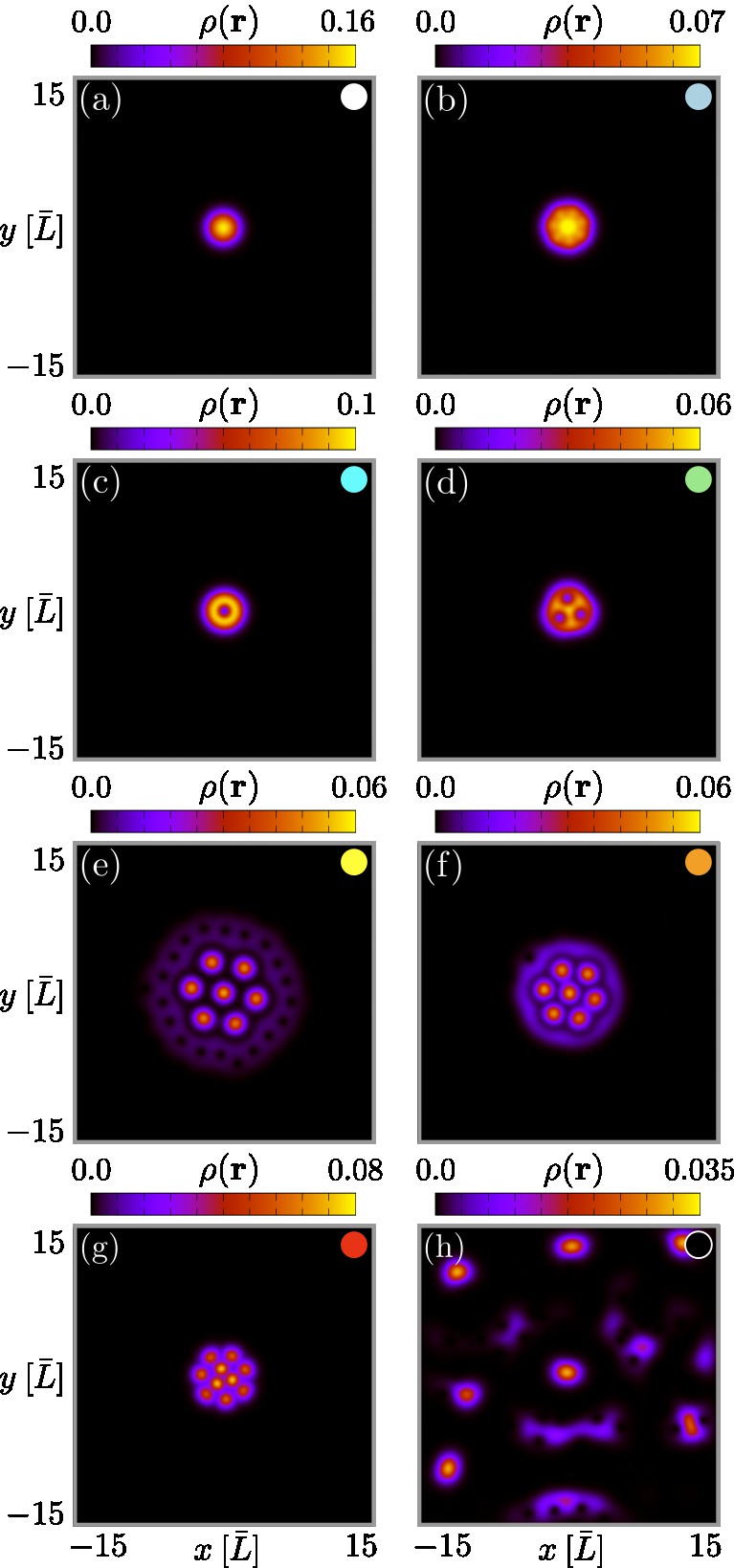}
\caption{
The different states appearing for $N=10$ dipolar bosons in a rotating trap:
(a) superfluid,
(b) fragmented superfluid,
(c) single-vortex threaded condensate,
(d) multiple-vortex threaded condensate,
(e) cluster lattice surrounded by vortices,
(f) cluster lattice surrounded by aura,
(g) single-particle clusters,
(h) disrupted condensate (rotational speed too high).
The circles in the upper-right corners have the same color code used in the phase diagram ~\ref{fig:PD-N-10} and indicate which states the density represents. 
}
\label{fig:density-N-10}
\end{figure}

From the phase diagram, we immediately see that the mean-field calculations [Fig.~\ref{fig:PD-N-10}(a)] give rise to a completely different picture than the calculations that account for quantum fluctuations [Fig.~\ref{fig:PD-N-10}(b)-(d)].
This is expected, since $N=10$ are typically not enough particles to form stable superfluids beyond weakly-interacting cases~\cite{Leggett:2001}.
In the mean-field phase diagram shown in Fig.~\ref{fig:PD-N-10}(a), only superfluid states and vortices appear.
These states have density profiles similar to the ones observed in systems with a larger number of particles, as illustrated in Fig.~\ref{fig:density-N-10}(a),(d).
Accounting for correlations with $M>1$, though, gives rise to a progressively much richer landscape of states.
The changes are particularly pronounced for fast rotations and strong interactions.
Even with $M=2$ orbitals, a coexistence of clusters and vortices appears [yellow region, Fig.~\ref{fig:PD-N-10}(b)].
Surprisingly, we find that the lattice formed by the clusters is not square, but triangular, as evidenced by the density in Fig.~\ref{fig:density-N-10}(e).

When even more orbitals are added, the region hosting a combination of clusters and vortices progressively shrinks.
Instead, the cluster states appear without vortices but surrounded by a superfluid ``aura", as illustrated in Fig.~\ref{fig:density-N-10}(f).
The vast majority of the strongly-interacting regime for $M=10$ is, however, occupied by clusters of localized single-particle peaks similar to what shown in Fig.~\ref{fig:density-N-10}(g).

%%%%%%%%%%%%%%%%%%%%%%%%%%%%%%%%%%%%%%%%%%%%
%%%%%%%%%%%%%%%%%%%%%%%%%%%%%%%%%%%%%%%%%%%%

\begin{figure}
\centering
\includegraphics[width=1.0\columnwidth]{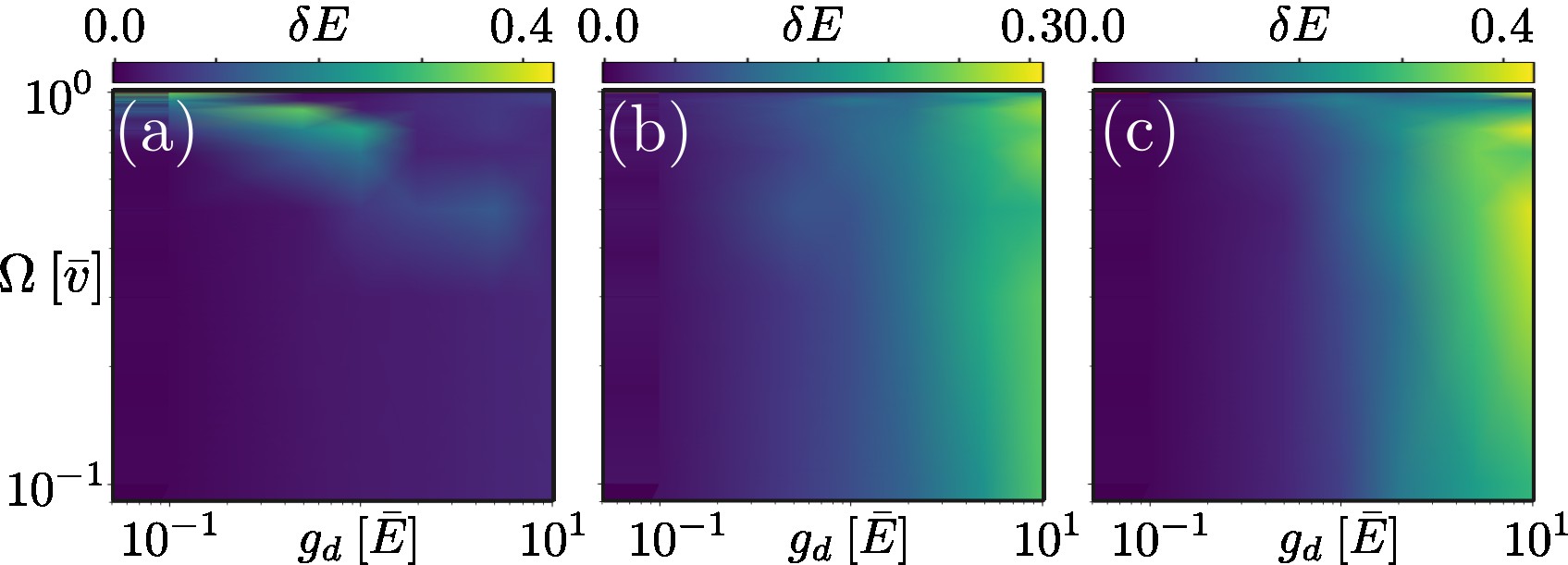}
\caption{
The difference in ground state energy between different $N=10$ calculations with increasing orbital number: (a) $\delta E = \frac{E_{M=1} - E_{M=2}}{E_{M=1}}$, (b) $\delta E = \frac{E_{M=2} - E_{M=5}}{E_{M=2}}$, (c) $\delta E = \frac{E_{M=5} - E_{M=10}}{E_{M=5}}$. 
}
\label{fig:energies-N-10}
\end{figure}

By examining the energy in Fig.~\ref{fig:energies-N-10}, we can appreciate that the calculations with a larger number of orbitals allow us to reach variational ground state with much lower energy.
Compared to the simulations with a larger number of particles, the decrease here is much more substantial, with peaks of up to -60\% in the strongly interacting regimes.
This indicates that the various kinds of cluster states obtained with MCTDH are strongly gapped from the mean-field states that one can extract from the GPE.

%%%%%%%%%%%%%%%%%%%%%%%%%%%%%%%%%%%%%%%%%%%%
%%%%%%%%%%%%%%%%%%%%%%%%%%%%%%%%%%%%%%%%%%%%

\begin{figure}
\centering
\includegraphics[width=1.0\columnwidth]{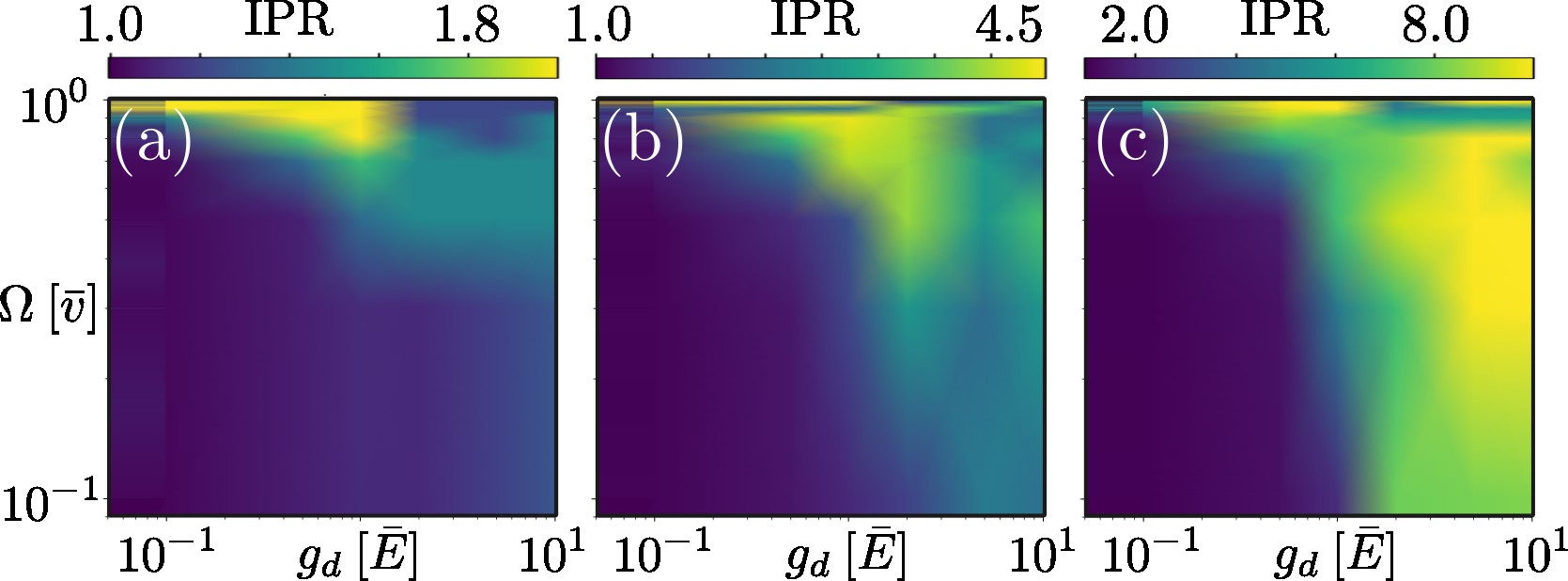}
\caption{
The IPR in parameter space for $N=10$ calculations with (a) $M=2$ and (b) $M=5$ orbitals, (c) $M=10$ orbitals.
}
\label{fig:ipr-N-10}
\end{figure}

Finally, examining the IPR in Fig.~\ref{fig:ipr-N-10} reveals that the average orbital occupation increases dramatically when more orbitals are included in the many-body state decomposition.
In fact, regimes of fast rotations and/or moderate to strong interactions saturate the variational space all the way to $M=10$, indicating the necessity of employing a large number of orbitals to describe the true ground state of the system.
Unfortunately, calculations with more than $M=10$ orbitals are currently beyond our computational capabilities.
Therefore, we cannot establish orbital convergence beyond this number.
As we will show in the next section, though, for a system of $N=5$ particles it is possible to claim a moderate level of orbital convergence for the cluster states described by $M=10$ orbitals.

%%%%%%%%%%%%%%%%%%%%%%%%%%%%%%%%%%%%%%%%%%%%
%%%%%%%%%%%%%%%%%%%%%%%%%%%%%%%%%%%%%%%%%%%%

\subsection{$N=5$ bosons}
\label{sec:N=5}

%%%%%%%%%%%%%%%%%%%%%%%%%%%%%%%%%%%%%%%%%%%%
\begin{figure}[t!]
\centering
\includegraphics[width=1.0\columnwidth]{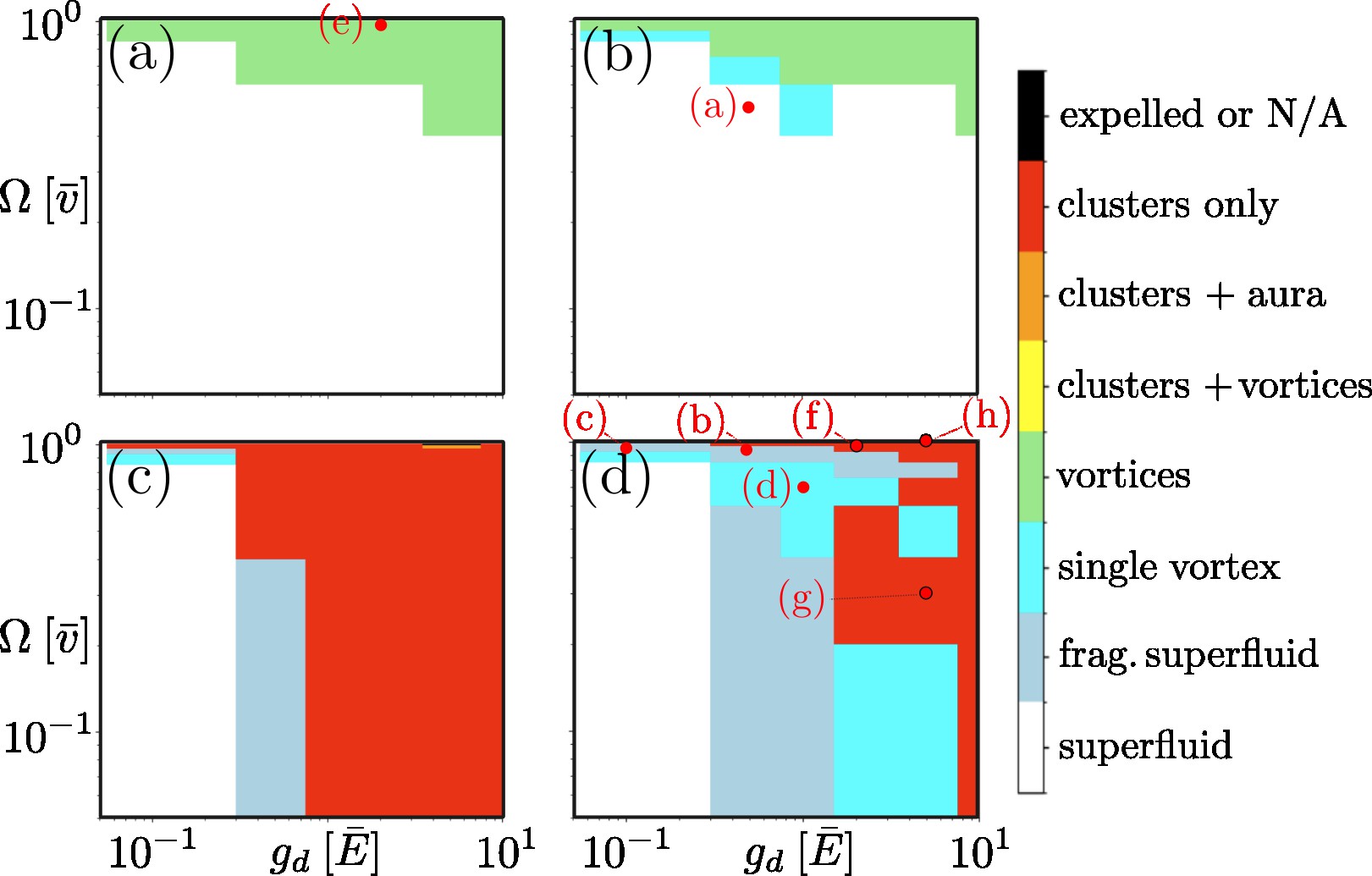}
\caption{
The different phases obtained from the simulation of $N=5$ dipolar-interacting bosons in a rotating potential as a function of interaction strength $g_d$ and rotation speed $\Omega$ (on a logarithmic scale).
The different panels show results obtained for different number of orbitals:
(a) $M=1$ orbitals,
(b) $M=2$ orbitals,
(c) $M=5$ orbitals.
(d) $M=10$ orbitals.
The red letters in the phase diagram refer to the points in parameter space where the density distributions of Fig.~\ref{fig:density-N-5} were probed.
}
\label{fig:PD-N-5}
\end{figure}
%%%%%%%%%%%%%%%%%%%%%%%%%%%%%%%%%%%%%%%%%%%%

\begin{figure}[t!]
\centering
\includegraphics[width=0.9\columnwidth]{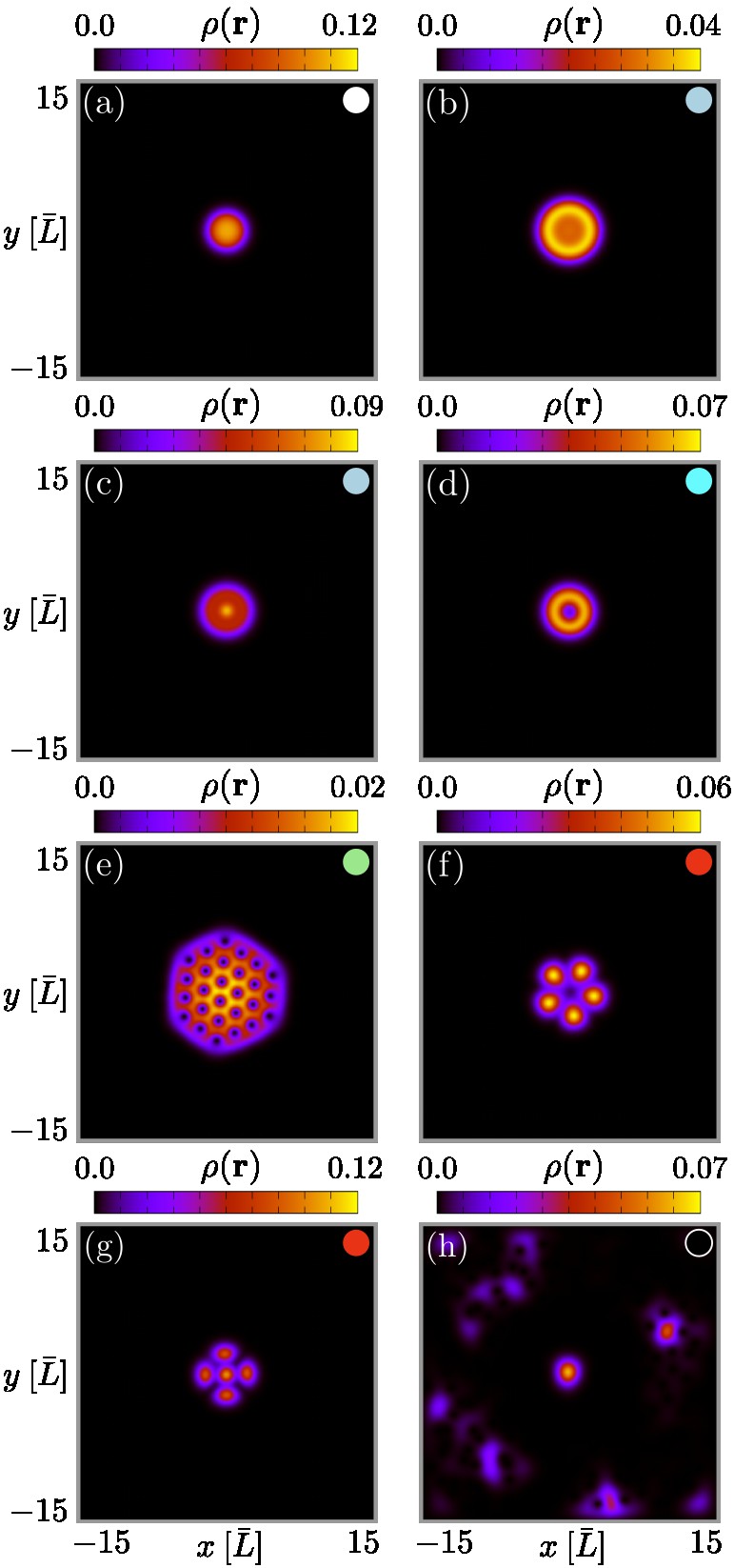}
\caption{
The different states appearing for $N=5$ dipolar bosons in a rotating trap:
(a) superfluid,
(b) fragmented superfluid (``blood cell" profile),
(c) fragmented superfluid (``droplet" profile),
(d) single-vortex threaded condensate,
(e) multiple-vortex threaded condensate (only appears for $M=1,2$),
(f) single-particle clusters (fivefold symmetric),
(g) single-particle clusters (fourfold symmetric),
(h) disrupted condensate (rotational speed too high).
The circles in the upper-right corners have the same color code used in the phase diagram ~\ref{fig:PD-N-5} and indicate which states the density represents. 
}
\label{fig:density-N-5}
\end{figure}

To confirm the convergence of our $N=10$ calculations, we have additionally performed simulations with just $N=5$ particles.
This few-body system should be indescribable with mean-field approximations.
Our results for the phase diagrams obtained with different numbers of orbitals are presented in Fig.~\ref{fig:PD-N-5}.
As observed already in the $N=10$ case, using a mean-field approximation (and even small deviations from it with two orbitals) leads to a parameter space dominated by superfluid states and vortices [cf. Fig.~\ref{fig:PD-N-5}(a)-(b)].
These states are fairly regular as shown in Fig.~\ref{fig:density-N-5}(a), (d), and (e).
In particular, the Abrikosov vortex lattice that arises at high rotation frequencies appears to have a very stable triangular structure.

%%%%%%%%%%%%%%%%%%%%%%%%%%%%%%%%%%%%%%%%%%%%
%%%%%%%%%%%%%%%%%%%%%%%%%%%%%%%%%%%%%%%%%%%%

However, these mean-field solutions are erroneous.
Only by sizably increasing the number of orbitals to $M=5$ [Fig.~\ref{fig:PD-N-5}(c)] and $M=10$ [Fig.~\ref{fig:PD-N-5}(d)] do the true ground states appear in the regimes of fast rotations and strong interactions.
These states consist mainly of exotic fragmented superfluids and cluster states.
The fragmented superfluid take ``blood cell" or droplet forms, as indicated in Fig.~\ref{fig:density-N-5}(b),(c).
We remark that these shapes were already found in other studies with a larger particle number~\cite{Schmidt:2021}.
Here, we discover that they persist for systems with as little as five bosons.
The cluster states appear in two different configurations with five-fold [Fig.~\ref{fig:density-N-5}(f)] or four-fold [Fig.~\ref{fig:density-N-5}(g)] symmetry.
We did not find a systematic criterion for when one or the other symmetry is preferred in our simulations, and thus suspect that these states are quasidegenerate.

\begin{figure}[t!]
\centering
\includegraphics[width=1.0\columnwidth]{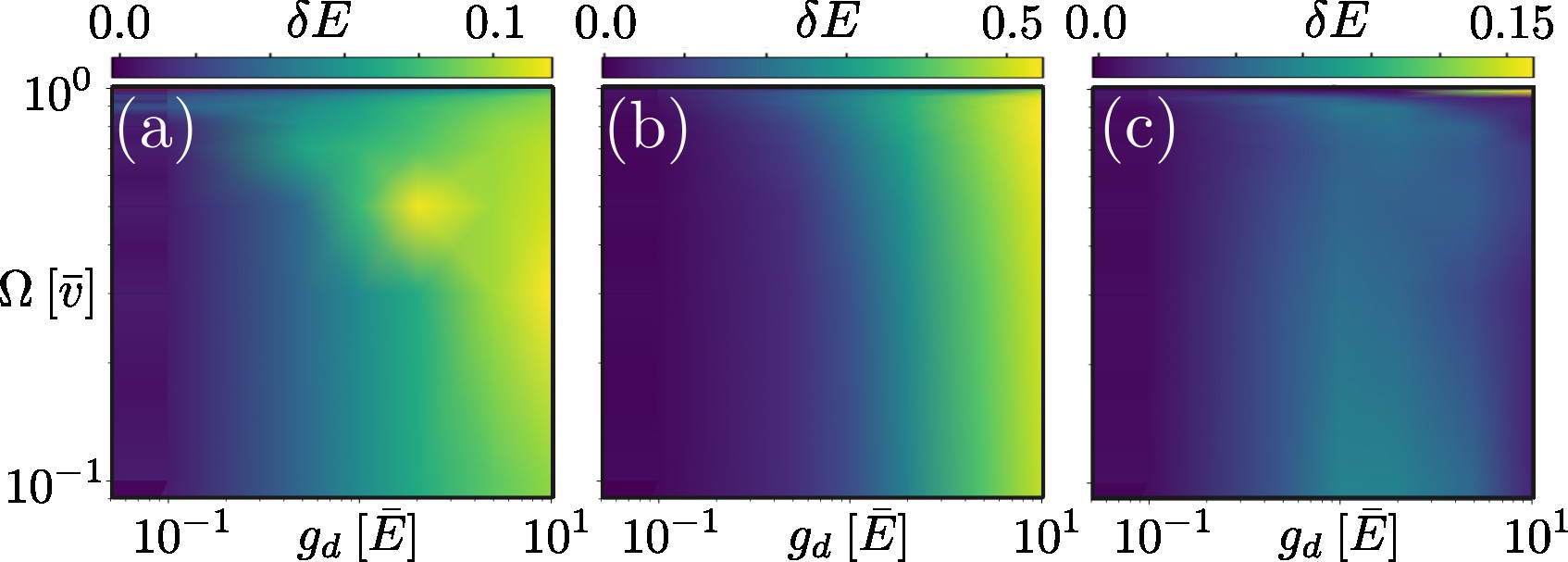}
\caption{
The difference in ground state energy between different $N=10$ calculations with increasing orbital number: (a) $\delta E = \frac{E_{M=1} - E_{M=2}}{E_{M=1}}$, (b) $\delta E = \frac{E_{M=2} - E_{M=5}}{E_{M=2}}$, (c) $\delta E = \frac{E_{M=5} - E_{M=10}}{E_{M=5}}$. 
}
\label{fig:energies-N-5}
\end{figure}

It is interesting to examine the energy and the IPR of the $N=5$ calculations as they contextualize better the convergence of our results with increasing orbital numbers.
The energy difference is plotted in Fig.~\ref{fig:energies-N-5}.
The energies of the multi-orbital calculations are systematically lower than the mean-field ones (with peaks of -80\% between $M=1$ and $M=10$).
Moreover, they decrease monotonically with an increasing number of orbitals, indicating progressive convergence.
Notably, the largest decrease in energy is observed when going from $M=2$ to $M=5$ calculations (around 50\% lower energy), while the step from $M=5$ to $M=10$ orbitals contributes an additional average 6\% (and maximum 15\%) to the total energy reduction.
Combined with a similar topology of the phase diagram and a saturation of the IPR (see below), this plateauing behavior suggests that the variational landscape with $M=10$ is large enough to accommodate the true ground state of the system.

\begin{figure}[t!]
\centering
\includegraphics[width=1.0\columnwidth]{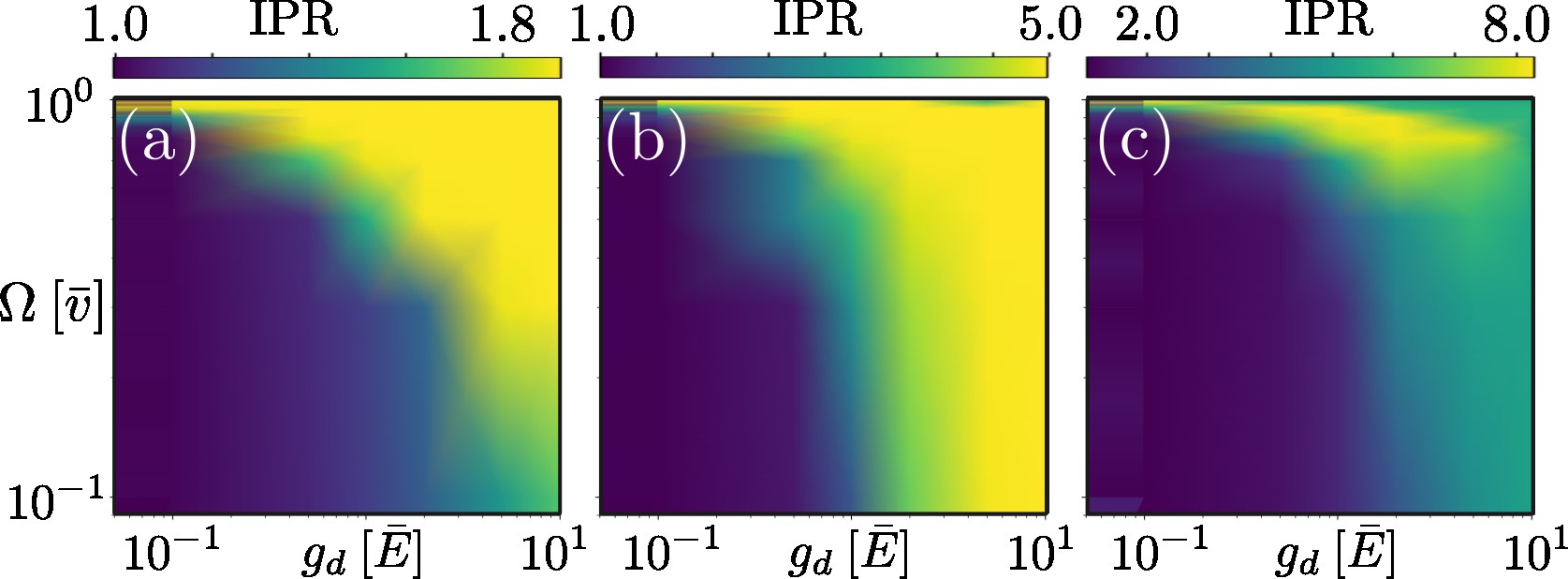}
\caption{
The IPR in parameter space for $N=5$ calculations with (a) $M=2$ and (b) $M=5$ orbitals, (c) $M=10$ orbitals.
}
\label{fig:ipr-N-5}
\end{figure}

The results pertaining to energy convergence are further corroborated by the IPR behavior, presented in Fig.~\ref{fig:ipr-N-5}.
In regimes of fast rotations and strong interactions, both the $M=2$ and $M=5$ calculations lead to a completely filled variational subspace. 
However, the vast majority of the parameter space for $M=10$ calculations yields an average orbital occupation of five [green color Fig.~\ref{fig:ipr-N-5}(c)], with peaks of eight for fast rotations.
This indicates a good degree of orbital convergence and validates the correctness of the variational cluster states representing the true ground states.

%%%%%%%%%%%%%%%%%%%%%%%%%%%%%%%%%%%%%%%%%%%%
%%%%%%%%%%%%%%%%%%%%%%%%%%%%%%%%%%%%%%%%%%%%
%%%%%%%%%%%%%%%%%%%%%%%%%%%%%%%%%%%%%%%%%%%%
%%%%%%%%%%%%%%%%%%%%%%%%%%%%%%%%%%%%%%%%%%%%

%%%%%%%%%%%%%%%%%%%%%%%
%%%%% CONCLUSIONS %%%%%
%%%%%%%%%%%%%%%%%%%%%%%
%%%%%%%%%%%%%%%%%%%%%%%%%%%%%%%%%%%%%%%%%%%%%%%%%%%%%%%%%%%%%%%%%%%%%%%%%%%%%%%%%%%%%%%%%%%%%
\section{Conclusions and Outlook} 
\label{sec:conclusions}

In this work, we have investigated the many-body ground-state properties of rotating dipolar Bose gases beyond the mean-field approximation. 
By employing the MultiConfigurational Time-Dependent Hartree (MCTDH) method for indistinguishable particles, we systematically explored the emergence of correlation-driven quantum phases in systems of finite-size condensates containing between $N=5$ and $N=50$ bosons. 
Our results highlight that mean-field descriptions fail to capture crucial quantum correlations that arise in regimes of fast rotation and strong dipolar interactions. 
More specifically, by including up to $M=10$ orbitals in our variational ansatz, we have mapped out a rich phase diagram that includes fragmented superfluids, vortex lattices, cluster states, and hybrid configurations featuring coexisting vortices and density modulations.

We demonstrated that increasing the number of orbitals in our variational calculations systematically reduces the ground-state energy, confirming that beyond-mean-field effects become significant in mesoscopic dipolar condensates. 
The inverse participation ratio (IPR) provided an additional quantification of wave function fragmentation.
Surprisingly, these effects are already visible for $N=50$ particles, with the appearance of roton instabilities in fragmented condensates, and their subsequent evolution into properly localized density clusters that can coexist with vortices.
In systems with lower particle number, we observed that the system transitions entirely from superfluid and vortex-dominated phases to strongly correlated cluster states when enough orbitals are taken into consideration, suggesting a fundamental breakdown of mean-field physics in small dipolar condensates.

Our work underscores the necessity of beyond-mean-field methods in capturing the intricate physics of rotating dipolar condensates and sets the stage for further theoretical and experimental explorations of these unconventional quantum phases in mesoscopic ultracold atomic systems. 
One immediate research direction to pursue is a detailed analysis of the correlation patterns in the identified phases, particularly focusing on the interplay between long-range interactions and vortex clustering~\cite{Prasad:2019, Klaus:2022, Bland:2023, Prasad:2024, Sabari:2024, Villois:2024}. 
A key open question is whether the cluster states observed in this study exhibit true supersolid ordering, characterized by simultaneous phase coherence and density modulation~\cite{Tanzi:2019, Tanzi:2019-2, Guo:2019, Chomaz:2019, Boettcher:2019, Poli:2024, Casotti:2024}. 
Further MCTDH investigations, or the use of other techniques such as quantum Monte Carlo~\cite{Pilati:2005, Buchler:2007, Macia:2014}, could help confirm the nature of these phases and their potential experimental realizability.

Another important aspect to consider is the exploration of how the full angular dependence of dipole-dipole interactions affects the stability and structure of the different phases~\cite{Lahaye:2009, Baranov:2012, Macia:2014, Lu:2015, Prasad:2021, Bland:2023}. 
In our study, we considered a simplified model where interactions are effectively isotropic and repulsive in the plane, which corresponds to dipoles placed perpendicularly to it.
A more complete treatment including anisotropic dipolar couplings could yield even richer phenomenology.
Additionally, studying the dynamical response of the different states we encountered to time-dependent protocols -- such as sudden quenches or periodic drives -- could provide insights into the nonequilibrium behavior of dipolar gases.
A particularly interesting direction involves continuous slowdowns, mimicking astrophysical processes like pulsar glitches.
Given that recent studies have linked pulsar glitches to rotating supersolids~\cite{Poli:2023, Bland:2024}, it is compelling to ask whether cluster states combined with vortices exhibit qualitatively distinct dynamical signatures in decelerating condensates.

Beyond bosonic systems, it would be fascinating to explore the fate of similar structures in fermionic gases subjected to rotation, particularly in regimes where the rotational speed approaches the Fermi velocity~\cite{Feder:2004, Zwierlein:2005, Tonini:2006, Cooper:2009, Baranov:2012, Kopyncinski:2021}. 
This could give rise to unconventional quantum Hall-like states or exotic topological phases stabilized by dipolar interactions~\cite{Mukherjee:2022, Predin:2023, Jiang:2024}. 

Lastly, an exciting prospect is extending our study to cavity-mediated long-range interactions, as encountered in condensates coupled to optical cavities~\cite{Ritsch:2013, Mivehvar:2021, Schlawin:2022}.
This setting introduces competition between vortex physics and superradiance, a problem that has so far been mostly addressed within mean-field approaches~\cite{Masalaeva:2024}. 
Understanding the impact of quantum fluctuations in such cavity-QED setups could reveal novel phases with strong light-matter coupling effects, offering connections between dipolar quantum gases and emergent photonic materials.

%%%%%%%%%%%%%%%%%%%%%%%%%%%%
%%%%% ACKNOWLEDGEMENTS %%%%%
%%%%%%%%%%%%%%%%%%%%%%%%%%%%
%%%%%%%%%%%%%%%%%%%%%%%%%%%%%%%%%%%%%%%%%%%%%%%%%%%%%%%%%%%%%%%%%%%%%%%%%%%%%%%%%%%%%%%%%%%%%
\textit{Acknowledgements --}
We thank Francesca Ferlaino, Elena Poli, Thomas Bland, Pramodh Seranath Yapa, and Leonardo Bellinato Giacomelli for useful discussions.
This work was partly supported by the Swedish Research Council (grants 2018-00313 and 2024-05213) and Knut and Alice Wallenberg Foundation (KAW) via the project Dynamic Quantum Matter (2019.0068).
Computation time at the High-Performance Computing Center Stuttgart (HLRS), on the Sunrise Compute Cluster of Stockholm University, and on the Euler cluster at the High-Performance Computing Center of ETH Zurich is gratefully acknowledged. 
%%%%%%%%%%%%%%%%%%%%%%%%%%%%%%%%%%%%%%%%%%%%%%%%%%%%%%%%%%%%%%%%%%%%%%%%%%%%%%%%%%%%%%%%%%%%%

%%%%%%%%%%%%%%%%%%%%%%
%%%%% APPENDICES %%%%%
%%%%%%%%%%%%%%%%%%%%%%
\appendix
%%%%%%%%%%%%%%%%%%%%%%%%%%%%%%%%%%%%%%%%%%%%%%%%%%%%%%%%%%%%%%%%%%%%%%%%%%%%%%%%%%%%%%%%%%%%%
\section{MCTDH-X}
\label{app:MCTDHX}
In this appendix, we provide a summary of the numerical method used to compute the ground states for the few-boson systems discussed in the main text. 
We employ the MultiConfigurational Time-Dependent Hartree (MCTDH) method for indistinguishable particles, coded in the MCTDH-X software~\cite{Alon:2008, Lode:2016, Fasshauer:2016, Lin:2020, Lode:2020, Molignini:2025-SciPost, MCTDHX}. 
This approach is tailored to solve the many-body Schr\"{o}dinger equation and is particularly suited for investigating the ground state properties and dynamics of interacting ultracold systems, including dipolar gases. 
The power of MCTDH-X is underscored by the fact that it has been successfully applied to a broad range of ultracold atomic systems from noninteracting Pauli crystals~\cite{Xiang:2023}, to short-range interacting systems~\cite{Roy:2018, Dutta:2019, Schaefer:2020, Lode:2021, Lode:2021-10, Debnath:2023, Roy:2023, Dutta:2023, Dutta:2024, Haldar:2024, Roy:2024-09, Roy:2024-11, Chatterjee:2024, Bhowmik:2025, Chakrabarti:2025-2}, to dipolar interacting atoms and molecules~\cite{Fischer:2015, Chatterjee:2018, Chatterjee:2019, Bera:2019, Bera:2019-symm, Chatterjee:2020, Roy:2022, Hughes:2023, Bilinskaya:2024, Molignini:2024, Molignini:2024-2, Roy:2024-annals, Roy:2024-epjp, Molignini:2025-quasicryst1, Molignini:2025-quasicryst2, Chakrabarti:2025}, to ultracold gases coupled with cavity fields~\cite{Lode:2017, Lode:2018, Molignini:2018, Lin:2019, Lin2:2019, Lin:2020-PRA, Lin:2021, Molignini:2022, Rosa-Medina:2022}.
In this appendix, we will explicitly focus on bosonic systems only (MCTDHB), although MCTDH-X has been applied to fermionic gases and spinor condensates too.

The starting point of the approach is the many-body time-dependent Schrödinger equation
\begin{equation}
    i \hbar \partial_t \left| \Psi(t) \right> = \mathcal{H}(t) \left| \Psi(t) \right>,
\end{equation}
with the many-body Hamiltonian for $N$ bosons at positions $\mathbf{r}_j$,
\begin{equation}
\mathcal{H}(t) = \sum_{j=1}^N \left[ T(\mathbf{r}_j) + V(\mathbf{r}_j; t) \right] + \sum_{j<k}^N W(\mathbf{r}_j -\mathbf{r}_k, t).
\label{eq:Ham}
\end{equation}
The single-particle Hamiltonian is composed of a kinetic part $T(\mathbf{r}_j)$ and a (potentially time-dependent) one-body potential $V(\mathbf{r}_j; t)$.
Moreover, the Hamiltonian contains (potentially time-dependent) two-body interactions $W(\mathbf{r}_j - \mathbf{r}_k; t)$.
In the present study, we consider dipolar bosons in a rotating harmonic trap, which can be described in the (static) rotating frame with kinetic energy that includes an angular contribution, $T(\mathbf{r}) = -\frac{\hbar^2}{2 m}\nabla_{\mathbf{r}}^2 - \Omega L_z$, a trap $V(\mathbf{r}) = \frac{1}{2} \omega \mathbf{r}^2$, and interactions $W(\mathbf{r}_j - \mathbf{r}_k) = \frac{g_d}{|\mathbf{r}_j - \mathbf{r}_k|^3 + \alpha}$, with $\alpha$ a regularization factor.
In the kinetic energy term, $L_z$ is the angular momentum stemming from the uniform rotation with frequency $\Omega$, i.e. $L_z = -i\hbar \left( x \frac{\partial}{\partial y} - y \frac{\partial}{\partial x} \right)$.

In the MCTDHB method, the many-body wave function for $N$ bosons is expanded as a time-dependent linear combination of permanents:
\begin{equation}
\left| \Psi(t) \right>= \sum_{\mathbf{n}}^{} C_{\mathbf{n}}(t)\vert \mathbf{n};t\rangle.
\label{many_body_wf}
\end{equation}
In turn, the permanents are built from $M$ single-particle orbitals, which are time-dependent functions.
In second quantization, the permanent construction takes the form
\begin{equation}
\vert \mathbf{n};t\rangle = \prod^M_{k=1}\left[ \frac{(\hat{b}_k^\dagger(t))^{n_k}}{\sqrt{n_k!}}\right] |0\rangle,
\label{many_body_wf_2}
\end{equation}
where $\mathbf{n}=(n_1,n_2,...,n_M)$ is a vector enumerating the number of bosons in each orbital, constrained by $\sum_{k=1}^M n_k=N$.
In this formula, the state $|0\rangle$ is the vacuum, while $\hat{b}_k^\dagger(t)$ creates at time $t$ a boson in the $k$-th orbital $\psi_k(x)$, defined as:
%%%
\begin{eqnarray}
	\hat{b}_k^\dagger(t)&=&\int \mathrm{d}x \: \psi^*_k(x;t)\hat{\Psi}^\dagger(x;t), \\
	\hat{\Psi}^\dagger(x;t)&=&\sum_{k=1}^M \hat{b}^\dagger_k(t)\psi_k(x;t). \label{eq:def_psi}
\end{eqnarray}
%%%

The number of orbitals governs both the accuracy and the convergence of the MCTDH expansion.
For $M=1$, the ansatz reduces to a mean-field Gross-Pitaevskii equation (GPE) with a single (global) wave function.
For $M>1$, many-body correlations can be captured effectively beyond mean-field approximations. 
Strictly speaking, the method is exact when $M \to \infty$.
However, very often it is possible to obtain \emph{numerically} exact results with a finite (and often small) number of $M$ when the many-body system is well-contained in the variational subspace spanned by the orbitals~\cite{Lode:2012, Fasshauer:2016}.
The number of permanents in the MCTDHB decomposition (i.e. the different configurations in our variational subspace) is what determines the computational complexity of the problem, and it scales combinatorially with the particle and orbital number: $ \left(\begin{array}{c} N+M-1 \\ N \end{array}\right)$. 
This scaling is very unfavorable when either $M$ or $N$ (or both) are very large, and therefore acts as a computational limit for the applicability of the method to arbitrary systems.
In practice, $M$ is chosen to balance precision requirements with computational feasibility.

To obtain the many-body state from the MCTDH decomposition, the Dirac-Frenkel time-dependent variational principle is applied to the Schrödinger equation written with the ansatz for the given many-body Hamiltonian~\cite{TDVM81}.
This leads to a set of coupled integro-differential equations in the variational parameters, namely both the expansion coefficients $C_\mathbf{n}(t)$ and the orbitals $\psi_i(x;t)$.
Solving these differential equations with optimized methods in real time allows us to compute the time evolution of a given initial state.
If they are solved in imaginary time instead, we can obtain a variational approximation to the ground state.
In this work, we focus on the latter approach.

Once the optimized many-body state has been calculated, it is rather straightforward to extract information about system. 
For example, the one-body reduced density matrix (1-RDM) is obtained as
%%%
\begin{eqnarray}
\rho^{(1)}(x,x') = \sum_{kq=1}^M \rho_{kq}\psi_k(x)\psi_q(x'),
\label{eq:red-dens-mat}
\end{eqnarray}
%%%
where
%%%
\begin{eqnarray}
\rho_{kq} = \begin{cases}
\sum_\mathbf{n} |C_\mathbf{n}|^2 n_k, \quad & k=q, \\
\sum_\mathbf{n} C_\mathbf{n}^* C_{\mathbf{n}^k_q} \sqrt{n_k(n_q+1)}, \quad & k\neq q, \\
\end{cases}
\end{eqnarray}
%%%
and the summation is over all possible configurations of $\mathbf{n}$. 
The quantity $\mathbf{n}^k_q$ represents the state where one boson is removed from orbital $q$ and added to orbital $k$. 
Higher-order reduced density matrices $\rho^{(n)}(x,x')$ can be calculated following a similar approach but tracing over multiple configurations.

Once $\rho^{(1)}(x,x')$ is known, the density distribution for the $N$ particles can be read out from its diagonal elements as
%%%
\begin{equation}
	\rho(x) = \rho^{(1)}(x,x)/N.
\end{equation}
%%%
From the 1-RDM, we can also obtain information about the orbital occupation by calculating its spectral decomposition
\begin{equation} 
\rho^{(1)}(\mathbf{x},\mathbf{x}') = \sum_i \rho_i \phi^{(\mathrm{NO}),*}_i(\mathbf{x}')\phi^{(\mathrm{NO})}_i(\mathbf{x}). 
\label{eq:RDM1-app} 
\end{equation} 
The eigenvalues of the 1-RDM are denoted as $\rho_i$ and are called orbital occupations.
They quantify the population of the corresponding eigenfunctions of the 1-RDM, which are termed natural orbitals.
Examining these two quantities provides information about particle occupation in different orbitals and their localization properties.

%%%%%%%%%%%%%%%%%%%%%%%%%%%%%%%%%%%%%%%%%%%%%%%%%%%%%%%%%%%%%%%%%%%%%%%%%%%%%%%%%%%%%%%%%%%%%

%%%%%%%%%%%%%%%%%%%%%%%%%%%%%%%%%%%%%%%%%%%%%%%%%%%%%%%%%%%%%%%%%%%%%%%%%%%%%%%%%%%%%%%%%%%%%
\section{Units and system parameters}

This appendix presents the parameters used in the simulations discussed in the main text.
We have performed simulations with a variable number of particles and orbitals.
The number of particles is varied between $N=5$ and $N=50$ and the number of orbitals is varied between $M=1$ (mean-field approximation) and $M=10$.

In our MCTDH-X simulations, we choose the units for the different quantities as follows.
Since the particles are trapped in a harmonic potential, its trapping frequency $\omega$ gives a natural scale to base our units off of.

The unit of length is conveniently set as $\bar{L} \equiv \sqrt{\hbar/(m\omega)}$.
We then run simulations with 256 grid points in both $x$ and $y$ directions, in an interval $x, y \in [-16 \bar{L}, 16 \bar{L}]$, giving a resolution of 0.125 $\bar{L}$.

In MCTDH-X, the unit of energy $\bar{E}$ is set in terms of the unit of length as $\bar{E} \equiv \frac{\hbar^2}{m \bar{L}^2}$.
By inserting our choice for the unit of length, we immediately see that the unit of energy corresponds to the quantized energy of the harmonic trap, i.e. $\bar{E} = \hbar \omega$.

The unit of time in our simulations is also defined from the unit of length.
More precisely, we set $\bar{t} \equiv \frac{m \hat{L}^2}{\hbar}$.
After using our definition of unit of length, we can see that the expression simplifies to $\bar{t}=\frac{1}{\omega}$.
In other words, the unit of time in our simulations is the inverse frequency of the trap.

Finally, the unit of speed is defined as the ratio of the unit of length and the unit of time, i.e.
$\bar{v} \equiv \frac{\bar{L}}{\bar{t}} = \sqrt{\frac{\hbar \omega}{m}}$.

Additionally, in our numerics we work in natural units, meaning that we set $\hbar=m=1$, where $m$ is the mass of our particles, throughout our computations.

Table ~\ref{table:pars} summarizes the parameter values used to generate the density plots in Figs.~2, 6, 10, 14.

\begin{table}
\centering
\begin{tabular}{||c|c|c|c||}
\hline \hline
& interaction $g_d [\bar{E}]$ & velocity $\Omega [\bar{v}]$ & orbitals $M$ \\
\hline \hline
$N=50$ (Fig. 2) & & & \\
\hline \hline
(a) & 1.0 & 0.3 & 1 \\
\hline
(b) & 2.0 & 0.0 & 5 \\
\hline
(c) & 0.5 & 0.3 & 5 \\
\hline
(d) & 2.0 & 0.3 & 5 \\
\hline
(e) & 2.0 & 0.7 & 1 \\
\hline
(f) & 0.5 & 0.99 & 5\\
\hline \hline
$N=20$ (Fig. 6) & & & \\
\hline \hline
(a) & 0.1 & 0.5 & 1 \\
\hline
(b) & 5.0 & 0.0 & 5 \\
\hline
(c) & 2.0 & 0.1 & 5 \\
\hline
(d) & 1.0 & 0.3 & 5 \\
\hline
(e) & 0.1 & 0.99 & 5 \\
\hline
(f) & 5.0 & 0.99 & 5 \\
\hline \hline
$N=10$ (Fig. 10) & & & \\
\hline \hline
(a) & 0.1 & 0.1 & 10 \\
\hline
(b) & 0.5 & 0.3 & 10 \\
\hline
(c) & 0.1 & 0.8 & 10 \\
\hline
(d) & 0.1 & 0.9 & 10 \\
\hline
(e) & 5.0 & 0.9 & 10 \\
\hline
(f) & 5.0 & 0.99 & 5 \\
\hline
(g) & 0.5 & 0.8 & 10 \\
\hline
(h) & 5.0 & 1.0 & 10 \\
\hline \hline
$N=5$ (Fig. 14) & & & \\
\hline \hline
(a) & 0.5 & 0.5 & 2 \\
\hline
(b) & 0.5 & 0.95 & 10 \\
\hline
(c) & 0.1 & 0.95 & 10 \\
\hline
(d) & 1.0 & 0.7 & 10 \\
\hline
(e) & 2.0 & 0.95 & 1 \\
\hline
(f) & 2.0 & 0.95 & 10 \\
\hline
(g) & 5.0 & 0.3 & 10 \\
\hline
(h) & 5.0 & 1.0 & 10 \\
\hline \hline
\end{tabular}
\caption{Parameters used to generate the density plots of Figs. 2, 6, 10, 14 in the main text.}
\label{table:pars}
\end{table}

%%%%%%%%%%%%%%%%%%%%%%%%%%%%%%%%%%%%%%%%%%%%%%%%%%%%%%%%%%%%%%%%%%%%%%%%%%%%%%%%%%%%%%%%%%%%%
\bibliography{biblio}
%%%%%%%%%%%%%%%%%%%%%%%%%%%%%%%%%%%%%%%%%%%%%%%%%%%%%%%%%%%%%%%%%%%%%%%%%%%%%%%%%%%%%%%%%%%%%

\end{document}